\documentclass[prc,aps,fonts,amsmath,superscriptaddress,floatfix,twocolumn]{revtex4-1}
\usepackage{graphicx,float} 
\usepackage{color}
\usepackage{mwe}
\usepackage{braket}
\usepackage{array}
\usepackage{soul}

\usepackage[caption=false]{subfig}
\usepackage{hyperref}
\newcolumntype{P}[1]{>{\centering\arraybackslash}p{#1}}

\begin{document}

\title{Low-energy monopole strength in spherical and deformed nuclei :  cluster and soft modes}

\author{F. Mercier}
\affiliation{IJCLab, Universit\'e Paris-Saclay, IN2P3-CNRS, F-91406 Orsay Cedex, France}

\author{J.-P. Ebran}
\affiliation{CEA,DAM,DIF, F-91297 Arpajon, France}
\affiliation{Universit\'e Paris-Saclay, CEA, Laboratoire Mati\`ere en Conditions Extr\^emes, 91680, Bruy\`eres-le-Ch\^atel, France}

\author{E. Khan}
\affiliation{IJCLab, Universit\'e Paris-Saclay, IN2P3-CNRS, F-91406 Orsay Cedex, France}

\begin{abstract} 
\begin{description}
\item[Background] Several recent experiments report significant low-energy isoscalar monopole strength, below the giant resonance, in various nuclei. In light $\alpha$-conjugate nuclei, these low-energy resonances were recently interpreted as cluster vibration modes. However, the nature of these excitations in neutron-rich nuclei remain elusive.
\item[Purpose] The present work provides a systematic analysis of the low-energy monopole strength in isotopic chains, from Neon to Germanium, in order to monitor and understand its nature and conditions of emergence.
\item[Methods] We perform covariant quasiparticle random phase approximation (QRPA) calculations, formulated within the finite amplitude method (FAM), on top of constrained relativistic Hartree-Bogoliubov (RHB) reference states.
\item[Results] Neutron excess leads to the appearance of low-energy excitations according to a systematic pattern reflecting the single-particle features of the underlying RHB reference state. With the onset of deformation, these low-energy resonances get split and give rise to more complex patterns, with possible mixing with the giant resonance. At lower energy, cluster-like excitations found in $N=Z$ systems survive in neutron-rich nuclei,  with valence neutrons arranging in molecular-like orbitals. Finally, at very low energy, pair excitations are also found in superfluid nuclei, but remain negligible in most of the cases.
\item[Conclusions] The low-energy part of the monopole strength exhibits various modes, from cluster vibrations ($\sim$ 5-10 MeV) to components of the giant resonance downshifted by the onset of deformation, including soft modes ($\sim$ 10-15 MeV) as well as pair excitation ($<$ 5 MeV), with possible mixing, depending on neutron-excess, deformation, and pairing energy.
\end{description}
\end{abstract}
 
%\date{\today}

\maketitle

%===================================================================================================================
\section{Introduction}
%===================================================================================================================
%

The complex nature of nuclear many-body systems is reflected in the vast diversity of their structure properties and excitation modes. In stable nuclei, the response to an external perturbation is dominated by collective modes, involving the coherent superposition of particle-hole excitations, coined giant resonances (GRs). The structural evolution of nuclei as one drifts away from the valley of $\beta$-stability --- e.g. the onset of static correlations responsible for nuclear deformation, superfluidity and clustering; the exotic arrangements of nucleons stemming from an increasing unbalance between the neutron and proton number; the impact of the loosely-bound nature of nucleon orbitals and proximity of the continuum, etc. --- yields an enrichment of the ways a nucleus responds to an external probe. Namely, in such cases, one observes a redistribution of the strength functions towards lower energy, below the GRs, associated with the emergence of new, exotic patterns of excitation \cite{paa07,sav13,bra19}. The nature of these new excitations is basically twofold, namely (i) so-called soft modes involving resonant oscillations of a neutron skin against a tightly bound core~\cite{kha11,pei14,pie17,gam19}, or non-resonant individual excitations~\cite{yuk13,ham14} and (ii) cluster vibrations~\cite{chi15,chi16,kan16,kan20,mer21}. 

Soft modes have attracted much attention both from the experimental~\cite{kuh81,sta82,bel82,han87,kob89,aum99,sav13,bra19} and theoretical perspectives, with various approaches used to pin down their properties, including the (quasi-particle) Random Phase Approximation ((Q)RPA)~\cite{vre01,ter05,paa09},  the (quasi-particle) Finite Amplitude Method ((Q)FAM)~\cite{ina11,pei14} and the Multiphonon Quasiparticle-Phonon Model~\cite{rye02,tso04,tso04_2,tso08}. Most of the studies on soft modes, focused on dipole resonances and/or spherical systems (see however Ref.~\cite{yos10} for a discussion on monopole soft modes in neutron-rich deformed nuclei). A comprehensive understanding of the evolution of the low-energy part of the isocalar monopole (ISM) strength, with isospin asymmetry and deformation, is still lacking.

On the other hand, cluster modes have mostly been investigated in light $N=Z$ nuclei, both within the Antisymmetrized Molecular Dynamics (AMD) and Energy Density Functional (EDF) approaches~\cite{chi15,chi16,kan16,kan20,mer21}. How neutron excess interferes with such cluster modes, remains an open question.
From a more general point of view, the interplay between soft, cluster and GR modes, and their evolution with isospin asymmetry and deformation, have not been established in a single approach, mainly because of the lack of microscopic theoretical framework capable of simultaneously tackling these various modes within a computationally affordable effort. 

Recently, relativistic EDFs were shown to consistently describe both liquid- and cluster-like features of nuclei, be it their ground-state~\cite{ebr12,ebr14,ebr14b,ebr18,ebr20} or spectroscopic (energies of excited states, reduced probability transition, elastic and inelastic form factors, etc.) properties ~\cite{mar18,mar19}. Since a covariant formulation of the quasiparticle random phase approximation (QRPA), is known to correctly describe GRs and soft modes~\cite{vre01,paa09,nak07,ina09,avo11,lia13,nik13,tom16,mus16} on one hand, as well as cluster vibrations~\cite{mer21} on the other, it shall be a tool of choice for achieving a global understanding of the mechanisms driving the emergence of these modes. In this study, the covariant QRPA is implemented under the form of the QFAM~\cite{nik13,bje20}, which significantly lowers the computational cost, compared to the traditional matrix formulation. This approach allows to provide an in-depth study of the impact of (i) isospin asymmetry, (ii) deformation, and (iii) pairing correlations, as well as their interplay on the structure of the ISM strength.

%Plan
The paper is organized as follows. In Sec.~\ref{sec:qfam}, we briefly introduce the covariant QFAM formalism. The evolution of the ISM strength with isospin asymmetry is extensively discussed, in the simple case of spherical nuclei, in Sec.~\ref{sec:asym}. Sec.~\ref{sec:def} is dedicated to the consequences of the onset of deformation, on the properties of the ISM strength.
Finally, in Sec.~\ref{sec:pairing}, we analyse the role played by pairing correlations, in driving the emergence of another type of low-energy resonance.

%===================================================================================================================
%
%===================================================================================================================
\section{\label{sec:qfam} QFAM theoretical framework}
%===================================================================================================================
%

The present implementation of the covariant QFAM is based on Ref.~\cite{bje20}. Refs \cite{sto11,kor15} provide a presentation of the QFAM in a non-relativistic context.
In the QFAM formalism, an external time-dependent field
\begin{equation}
F(t) = \eta \left(F(\omega) e^{-i\omega t} + F^\dagger(\omega) e^{+i\omega t}   \right),
\label{Eq:linearization}
\end{equation} 
with $\eta$ a real, small parameter, induces a linear response of the system,
characterized by the following equations, in the quasiparticle (qp) basis:
\begin{subequations}
\begin{align}
\label{Eq:QFAM-20}
(E_\mu + E_\nu - \omega) X_{\mu \nu}(\omega) + \delta H_{\mu \nu}^{20}(\omega)  &= -F_{\mu \nu}^{20},\\
\label{Eq:QFAM-02}
(E_\mu + E_\nu + \omega) Y_{\mu \nu}(\omega) + \delta H_{\mu \nu}^{02}(\omega)  &= -F_{\mu \nu}^{02},
\end{align}
\end{subequations}
They describe the oscillation of the system around a static configuration, which is the solution of a constrained Relativistic Hartree-Bogoliuvob (RHB) equation
\begin{equation}
\begin{pmatrix}
    h(\boldsymbol{q})-\lambda & \Delta(\boldsymbol{q}) \\
    -\Delta^*(\boldsymbol{q}) & -h^*(\boldsymbol{q}) + \lambda
    \end{pmatrix}\begin{pmatrix}
    U_\mu(\boldsymbol{q}) \\
    V_\mu(\boldsymbol{q})
    \end{pmatrix} = E_\mu(\boldsymbol{q}) \begin{pmatrix}
    U_\mu(\boldsymbol{q}) \\
    V_\mu(\boldsymbol{q})
    \end{pmatrix}.  
\end{equation}
$E_\mu$, $U_\mu$ and $V_\mu$ stand for the energy and wavefunction of the qp $\mu$. The fields $h[\rho]$ and $\Delta[\kappa]$, functionals of the one-body normal and anomalous density matrices $\rho = V^*V^T$ and $\kappa = V^*U^T$, are the RHB mean potential in the particle-hole and particle-particle channels respectively. $\lambda$ is the chemical potential, $\boldsymbol{q}$ collects a set of constrained collective coordinates (e.g. deformation parameters, pairing gap, etc.). $X_{\mu \nu}(\omega)$ and 
$Y_{\mu \nu}(\omega)$ are the QFAM amplitudes at a given excitation energy $\omega$,  
$\delta H^{20(02)}$ ($F^{02(02)}$) represent the two-qp components of the induced Hamiltonien (external perturbation). Namely, if $F(\omega)$ is a one-body operator, represented by the matrix elements $f$,
\begin{equation}
    F(\omega) = \sum_{ij}f_{ij}c^\dagger_i c_j = \frac{1}{2} \begin{pmatrix}
    c^\dagger & c
    \end{pmatrix} \begin{pmatrix}
    f & 0 \\
    0 & -f^T
    \end{pmatrix}  \begin{pmatrix}
    c \\
    c^\dagger
    \end{pmatrix},
\end{equation}
(where a constant term was neglected) then
\begin{align}
&\begin{pmatrix}
    F^{11} & F^{20} \\
    F^{02} & -(F^{11})^T
    \end{pmatrix} \equiv \mathcal{W}^\dagger \begin{pmatrix}
    f & 0 \\
    0 & -f^T
    \end{pmatrix}\mathcal{W} \nonumber \\
    &\phantom{space}=\begin{pmatrix}
    U^\dagger f U - V^\dagger f^T V & U^\dagger f V^* - V^\dagger f^T U^* \\
    V^T f U - U^T f^T V & V^T f V^* - U^T f^T U^*
    \end{pmatrix},
\end{align}
where $\mathcal{W}$ is the unitary Bogoliubov transformation, by which the qp ladder operators $\beta^\dagger$ and $\beta$ are expressed as linear combinations of the single-particle (sp) operators $c$ and $c^\dagger$:
\begin{equation}
\label{eq:bogotrans}
    \begin{pmatrix}
\beta \\
\beta^\dagger
\end{pmatrix}
\begin{pmatrix}
    U^\dagger & V^\dagger \\
    V^T & U^T
    \end{pmatrix}
\begin{pmatrix}
c \\
c^\dagger
\end{pmatrix} \equiv \mathcal{W}^\dagger \begin{pmatrix}
c \\
c^\dagger
\end{pmatrix}.
\end{equation}
Likewise, by denoting, in the sp basis, the induced fields in the particle-hole and particle-particle channels by $\delta h(\omega)$ and $\delta \Delta^{(\pm)}(\omega)$ respectively, we have  
\begin{equation}
\begin{pmatrix}
    \delta H^{11} & \delta H^{20} \\
    -\delta H^{02} & -(\delta H^{11})^T
    \end{pmatrix}\equiv \mathcal{W}^\dagger \begin{pmatrix}
    \delta h & \delta\Delta^{(+)} \\
    -\delta\Delta^{(-)*} & -\delta h^T 
    \end{pmatrix} \mathcal{W},
\end{equation}
i.e.
\begin{align}
\label{eq:H20}
 \delta H^{20}(\omega) = &U^\dagger \delta h(\omega) V^* + U^\dagger \delta\Delta^{(+)}(\omega) U^* \nonumber \\
 &-V^\dagger \delta\Delta^{(-)*}(\omega) V^* - V^\dagger \delta h^T(\omega)  U^*, 
 \end{align}
 and
 \begin{align}
 \label{eq:H02}
  \delta H^{02}(\omega) = &-V^T \delta h(\omega) U - V^T \delta\Delta^{(+)}(\omega) V \nonumber \\
  &+ U^T \delta\Delta^{(-)*}(\omega) U + U^T \delta h^T(\omega)  V.
 \end{align}
The perturbed fields $\delta h$ and $\delta \Delta^{(\pm)}$ can be expressed in terms of the static RHB fields $h$ and $\Delta$ (after linearizing the explicit density-dependent parts of the covariant energy density functional) and the induced normal and anomalous density matrices, that is
\begin{subequations}
\begin{align}
\label{eq:perth}
    \delta h(\omega) &= h\left[\delta\rho(\omega)\right], \\
    \label{eq:pertd+}
    \delta \Delta^{(+)}(\omega) &= \Delta\left[\delta\kappa^{(+)}(\omega)\right],  \\
    \label{eq:pertd-}
    \delta \Delta^{(-)}(\omega) &= \Delta\left[\delta\kappa^{(-)}(\omega)\right],
\end{align}
\end{subequations}
and
\begin{subequations}
\begin{align}
\label{eq:pertrho}
\delta\rho(\omega) &= U X(\omega) V^T + V^* Y^T(\omega) U^\dagger, \\   
\label{eq:pertkap+}
\delta\kappa^{(+)}(\omega) &= UX(\omega)U^T + V^*Y^T(\omega)V^\dagger, \\
\label{eq:pertkap-}
\delta\kappa^{(-)}(\omega) &= UY^*(\omega)U^T + V^*X^\dagger(\omega) V^\dagger.
\end{align}
\end{subequations}
 The latter depend on the QFAM amplitudes $X$ and $Y$, making the master equations~\eqref{eq:H20} and~\eqref{eq:H02} self-consistent.

In the QFAM formalism, the strength function derives from 
\begin{equation}
S(f,\omega) = -\frac{1}{\pi} \textnormal{Im} \textnormal{Tr} \left[ f^\dagger \delta \rho(\omega) \right],
\label{eq:SF}
\end{equation}
The electric isoscalar multipole operator reads 
\begin{equation}
 f_{JK}^{IS} = \sum_{i=1}^A{f_{JK}(\mathbf{r}_i)},
\end{equation}
with $f_{JK}(\mathbf{r}) = r^J Y_{JK}(\theta,\phi)$. In the case of the monopole mode, $f_{00}(\mathbf{r}) = r^2$. For an even-even axially symmetric nucleus, the operators $f_{JK}$ and $f_{J-K}$ yield identical strength functions. QFAM calculations can therefore be simplified, by using the operator $f^{(+)}_{JK} = \left( f_{JK} + (-1)^Kf_{J-K} \right)/\sqrt{2+2\delta_{K0}}$, and assuming $K\ge 0$.

%............................................................
\begin{figure*}%[!htb]
\centering
\includegraphics[width=1\textwidth]{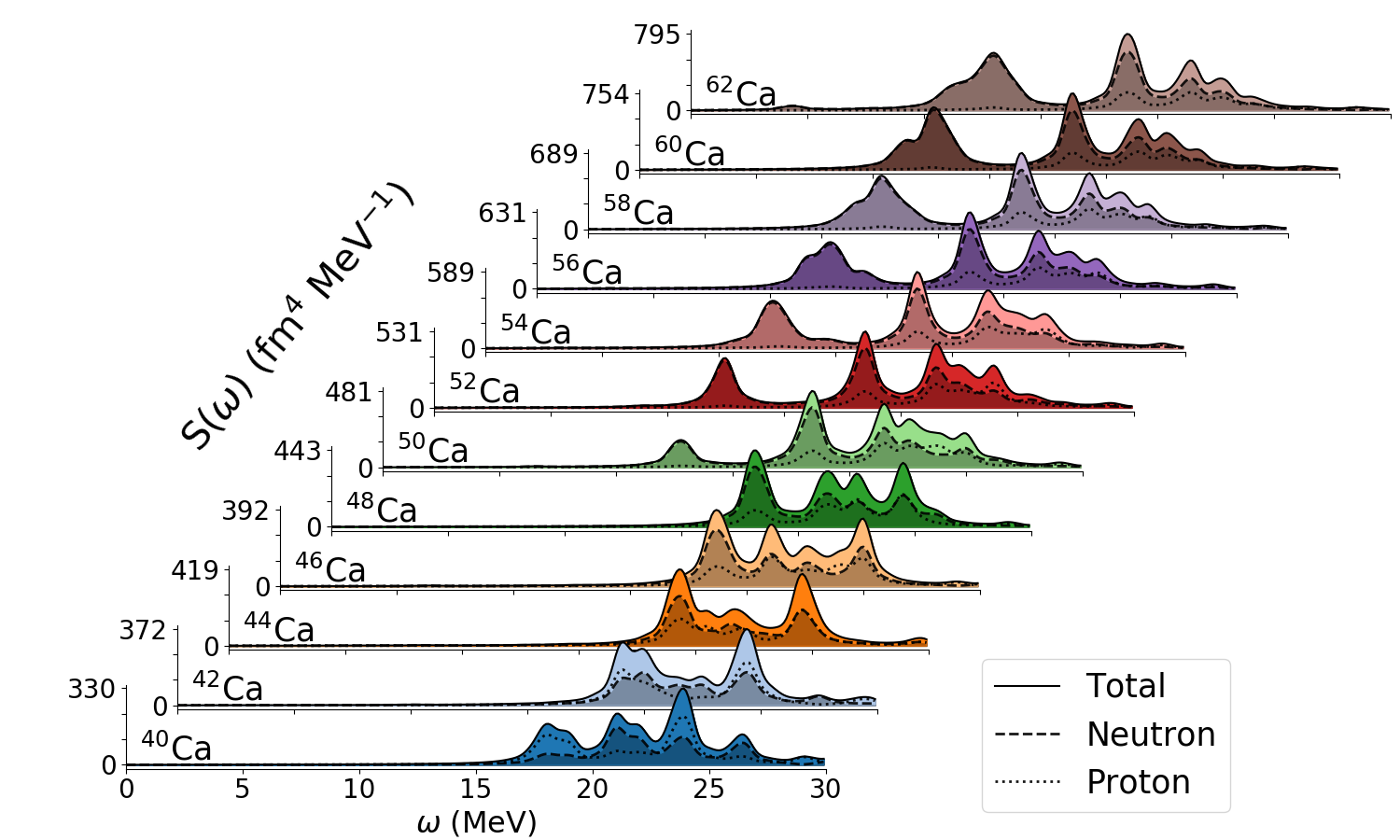}
%\par\end{centering}
\caption{Distribution of the isoscalar monopole strength in the Ca isotopic chain obtained within the covariant QFAM with the DD-PC1 parametrization, split into total (solid line), neutron (dashed line) and proton (dotted line) contributions.}
\label{fig:Ca_ISM_GS}
\end{figure*}
%............................................................

In this study, the QFAM calculations are based on the DD-PC1 energy density functional~\cite{nik08}, complemented
by a separable pairing force in momentum space~\cite{dug04,tia09}: 
$\displaystyle \langle k | V^{^1S_0}|k^\prime\rangle = -G p(k)p(k^\prime)$ in the particle-particle channel. By assuming a simple Gaussian ansatz
$p(k) = e^{-a^2k^2}$, the two parameters $G$ and $a$
were adjusted to reproduce the density dependence of the pairing gap
at the Fermi surface, obtained in nuclear matter with the Gogny D1S parametrization~\cite{ber91}.    
In practice, we first compute the $U$ and $V$ RHB matrices and qp energies $E_\mu$, after solving 
the RHB equations for a nucleus of interest, expanded in 
an axially symmetric harmonic oscillator basis. The QFAM procedure then consists 
of starting with a trial set of $X$'s and $Y$'s,
computing the induced density matrices~(\ref{eq:pertrho}-\ref{eq:pertkap-}),
then the perturbed mean fields~(\ref{eq:perth}-\ref{eq:pertd-}),    
and finally $H^{20}$ and $H^{02}$, according to Eqs.~\eqref{eq:H20}-\eqref{eq:H02}.
Solving~\eqref{Eq:QFAM-20}-\eqref{Eq:QFAM-02} then yields a new set of $X$'s and $Y$'s, from which the previous steps are repeated until convergence. Further details on the QFAM solver DIRQFAM can be found in Ref.~\cite{bje20}.

The obtained QFAM amplitudes (Eqs.~\eqref{Eq:QFAM-20}-\eqref{Eq:QFAM-02}), and the strength function~\eqref{eq:SF}, are defined in the whole complex $\omega$-plane, except at the 
QRPA eigenenergies, where they diverge. In practice, calculations are performed 
for excitation energies $\omega + i \gamma$, with a fixed imaginary part $\gamma$: this corresponds to a Lorentzian smearing of the strength function, with the width $\Gamma = 2\gamma$~\cite{avo11}. The smearing is fixed to $\Gamma=0.5$ MeV for all the calculations.
The size of the basis, in which the RHB-QFAM equations are expanded, runs from 13 to 15 shells, depending on the mass of the nucleus. It ensures the convergence of our results below 1\%. As explained in~\cite{mer21}, the part of the strength with $\omega > 20$ MeV may not be stable, with respect to the size of the basis. However, the position of the GMR centroids remains stable, and we aim to focus on the low-energy part of the strength.

%
%===================================================================================================================
\section{\label{sec:asym} Evolution of low-energy isoscalar monopole modes with isospin asymmetry }
%===================================================================================================================
%

We start by investigating the impact of the isospin asymmetry, on the properties of low-energy ISM modes by first focusing on the single open-shell $Z=20$ and $Z=28$ nuclei: the vanishing of deformation, at the mean field level, simplifies the pattern of excitation. Moreover, the abundance of both experimental and theoretical results, shall enable to benchmark the present calculations.

A generic pattern of emergence of low-energy ISM modes, as the neutron over proton numbers ratio rises, can be traced back to the mismatch between the neutron and proton Fermi energies, due to increasingly bound protons and conversely, last occupied neutron orbitals getting closer to the continuum. As a result, valence neutrons decouple from the other nucleons, and participate to excited modes of rather non-collective nature. In other words, the emergence of new low-energy ISM modes, with increasing neutron number, can be understood from the single-particle (sp) structure i.e. the sp spectrum of the reference RHB state, on top of which the QRPA response is built: the appearance of a peak, in the low-energy strength, coincides with the filling of an orbital with spherical quantum numbers $n,j,l,m$, from which an additional 2qp configuration $\left[njlm; (n+1)jlm\right]^{J=0}$  becomes available. The ISM strength therefore reflects the energy pattern of the sp spectrum: i) appearance of low-energy resonances, on top of the main part of the strength, whenever an orbital inaugurating a new major shell starts to get filled, and ii) increase of the strength in a small energy window, as long as orbitals of the same major shell are getting filled. The decrease of the corresponding excitation energy, as one goes from a major shell to the next one, mainly stems from the weakening of the binding energies of valence neutrons, i.e the shrinking of the gap between the occupied $n$ and empty $n+1$ orbitals, involved in the ISM transition. This specific pattern shall be illustrated on the $Z=20$ and $Z=28$ isotopic chains, in the next two Subsections.

\subsection{Calcium isotopes}

\begin{figure}[!htb]
\centering
\includegraphics[width=1\linewidth]{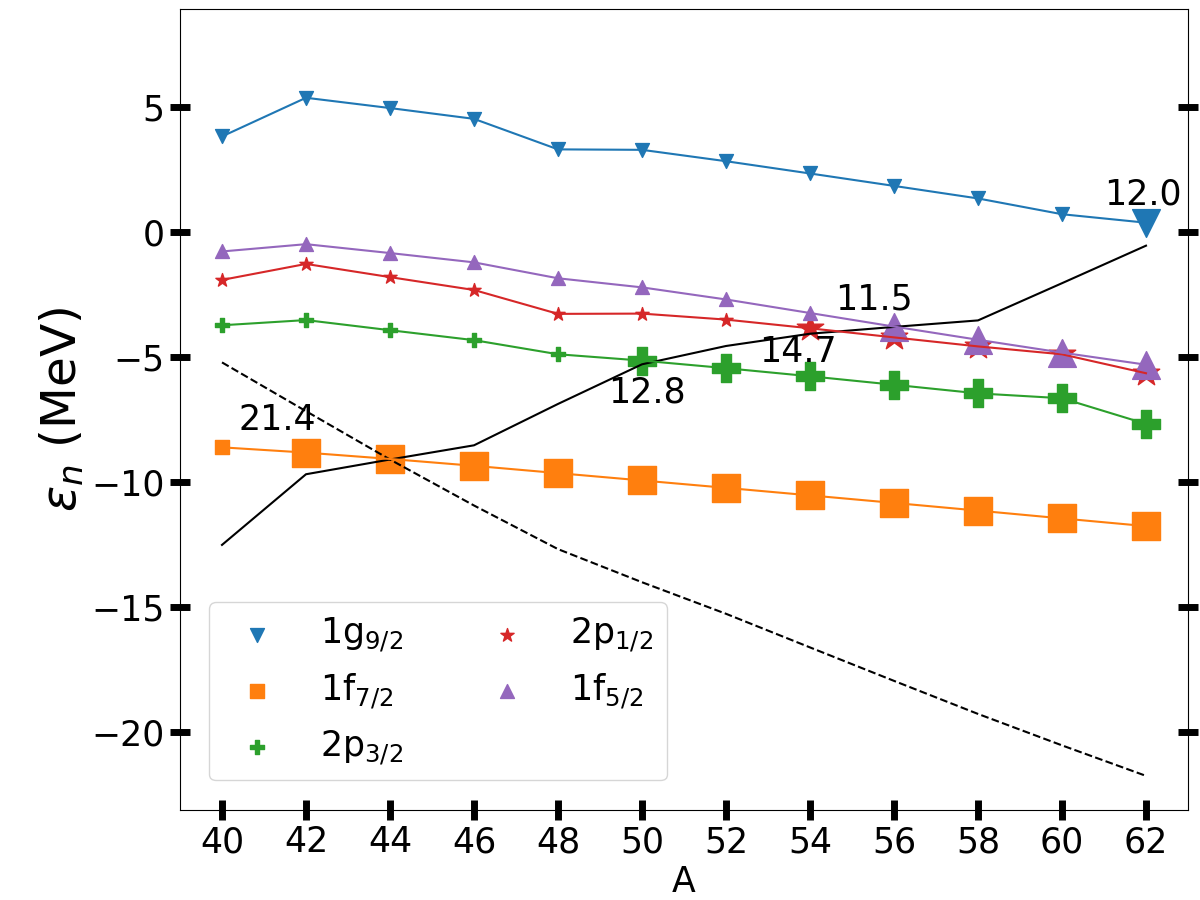}
%\par\end{centering}
\caption{RHB single-particle spectrum in Ca isotopes. The neutron (proton) Fermi energy is represented by a black solid (dashed) line. Empty (occupied) orbitals are designated by small (large) markers. Numbers indicate the energy of the resonance associated to the filled orbital.
}
\label{fig:SP_Ca}
\end{figure}

The distribution of ISM strength computed with the covariant QFAM in the even-even $^{40-62}$Ca isotopes is displayed in Fig.~\ref{fig:Ca_ISM_GS}. While the GMR ($\omega \sim 18$ MeV) remains quite stable along the isotopic chain, the structure of the low-energy part of the ISM strength changes with neutron excess. This can be understood with the Ca canonical single-neutron spectra, plotted against mass number, in Fig.~\ref{fig:SP_Ca}: in $^{42}$Ca, two neutrons fill the $1f_{7/2}$ orbital, opening the $1f_{7/2} \rightarrow 2f_{7/2}$ transition 
%(these two levels are separated by a $\sim$ 30 MeV energy gap) 
as a possible contribution to the ISM response. A decomposition of $^{42}$Ca monopole resonances into 2-qp components, shows that the $1f_{7/2} \rightarrow 2f_{7/2}$ transition mainly contributes to a peak located at $\omega \sim 20$ MeV, that is in the GMR. As discussed above, going to the next major shell, i.e. adding two neutrons in the $2p_{3/2}$ orbital, after the filling the $1f_{7/2}$ orbital --- that is $^{50}$Ca ---, is expected to generate a low-energy resonance, separated from the main part of the ISM strength. As also discussed above, filling the orbitals inside the same major shell, i.e from the $2p_{3/2}$ to the $1g_{9/2}$ levels, is expected to generate contributions to the strength in a same energy window. Indeed, the filling of each of the orbitals, from $2p_{3/2}$ to $1g_{9/2}$, gives rise to excitations close in energy (see Fig.~\ref{fig:SP_Ca}), each of them dominated by a single 2qp configuration. Namely, in $^{50}$Ca, one finds a 12.8 MeV excitation mode (see Fig.\ref{fig:Ca_ISM_GS} and Tab.~\ref{tab:Ca_qp}) for which the dominant 2qp contribution (carrying 40\% of the total 2qp contributions) is $2p_{3/2}\rightarrow 3p_{3/2}$, in agreement with the results of Refs~\cite{ham97,pie17,gam19}. In $^{54}$Ca, a small peak appears at 14.7 MeV, corresponding to the $2p_{1/2}\rightarrow 3p_{1/2}$ transition. In $^{56}$Ca, a resonance at 11.5 MeV emerges from the filling of the $1f_{5/2}$ orbital and the $1f_{5/2}\rightarrow 2f_{5/2}$ transition. In $^{62}$Ca, the filling of the $1g_{9/2}$, and hence the $1g_{9/2}\rightarrow 2g_{9/2}$ transition, triggers a resonance at 12 MeV. This last excitation is however not dominant, and is not clearly visible in the strength. However, the details of the qp contributions indicate that this transition holds for about 10\% of the strength between 10 and 13 MeV. It should be noted that another resonance is visible around 4 MeV in $^{62}$Ca, whose nature will be discussed in Sec.~\ref{sec:pairing}.
%............................................................
%............................................................

%............................................................
\begin{figure}[!htb]
\centering
\includegraphics[width=1\linewidth]{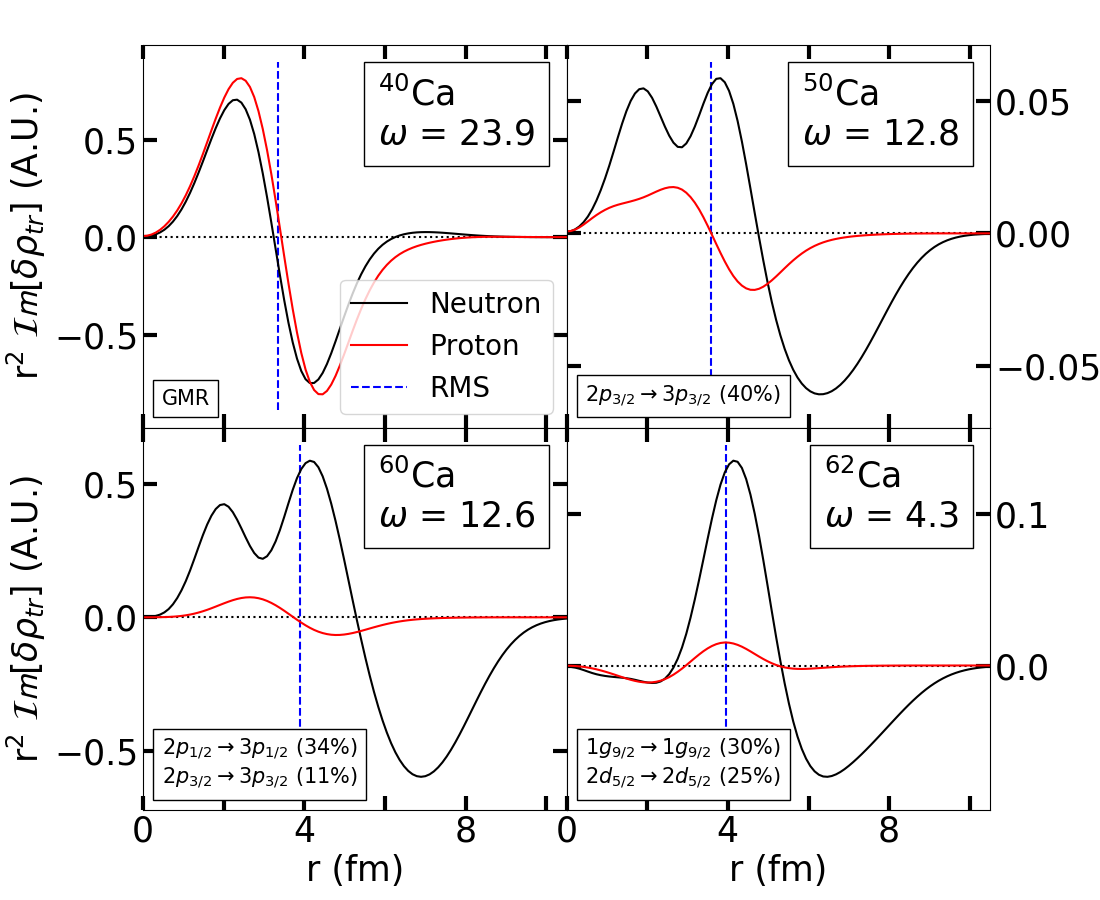}
%\par\end{centering}
\caption{ Neutron (black) and proton (red) transition densities for $^{40,50,60,62}$Ca. Excitation energies $\omega$ are given in MeV. The vertical blue line shows the ground-state r.m.s. matter radius of the nucleus.}
\label{fig:Ca_td}
\end{figure}
%............................................................

%............................................................
\begin{table}
%\begin{tabular}{ |p{1cm} | p{1cm} | p{2.5cm} | p{1.5cm} | p{1.5cm}|  }
\begin{tabular}{ |P{12mm}|P{12mm}|P{22mm}|P{15mm}|P{15mm}|  }
 \hline
 Nucleus & $\omega$ (MeV) & Transition $\alpha \rightarrow \beta$ & $E_\beta + E_\alpha$ (MeV) & $\left|X_{\alpha \beta}\right|^{2}-\left|Y_{\alpha \beta}\right|^{2}$  \\
 \hline \hline
 
 $^{50}$Ca & 12.8 & $2p_{3/2} \rightarrow 3p_{3/2}$ & 14.2 & 0.4  \\
 \hline
  
 $^{60}$Ca & 12.6 & $2p_{1/2} \rightarrow 3p_{1/2}$ & 12.4 & 0.34 \\
 & & $2p_{3/2} \rightarrow 3p_{3/2}$ & 13.8 & 0.11 \\
 & & $1f_{5/2} \rightarrow 2f_{5/2}$ & 14.2 & 0.05 \\
 \hline
 
 $^{62}$Ca & 4.3 & $1g_{9/2} \rightarrow 1g_{9/2}$ & 2.6 & 0.30 \\
 & & $2d_{5/2} \rightarrow 2d_{5/2}$ & 5.4 & 0.25 \\
 & & $2s_{1/2} \rightarrow 2s_{1/2}$ & 5.3 & 0.19 \\
 
 \hline

\end{tabular}
 \caption{Properties of the monopole modes found in Ca isotopes at excitation energy $\omega$. The 2qp decomposition (column transitions $\alpha\rightarrow\beta$), contribution to the strength $\left|X_{\alpha \beta}\right|^{2}-\left|Y_{\alpha \beta}\right|^{2}$ and  unperturbed energy $E_\beta+E_\alpha$ with $E_\mu$ the energy of the quasi-particle state $\mu$, are displayed. Transitions contributing less than 5\% to the ISM strength  are not reported. }
 \label{tab:Ca_qp}
\end{table}
%............................................................

The properties of the low-energy neutron modes can be further studied by computing their corresponding transition densities, which are plotted on Fig.~\ref{fig:Ca_td}. For the sake of comparison, the upper left panel displays the transition densities of the GMR in $^{40}$Ca, with its typical in-phase oscillation of protons and neutrons. The properties of the different modes are detailed in Tab.\ref{tab:Ca_qp}.
The transition densities associated to the low-lying modes at $\omega=12.8$ MeV in $^{50}$Ca and at $\omega=12.6$ MeV in $^{60}$Ca, show a very different behavior, compared to the GMR in $^{40}$Ca. As described in Ref.~\cite{gam19},  the valence neutrons seem to be decoupled from the protons, with neutron excitations extending over the whole volume of the nucleus. The proton contribution decreases with neutron excess, going from $~15$\% for $^{50}$Ca, to less than $~5$\% in the case of $^{62}$Ca.
The 4.3 MeV mode in $^{62}$Ca, driven by pairing correlations, will be discussed in Sec.~\ref{sec:pairing}.

\subsection{Nickel isotopes}

%............................................................
\begin{figure*}[!htb]
\centering
\includegraphics[width=1\linewidth]{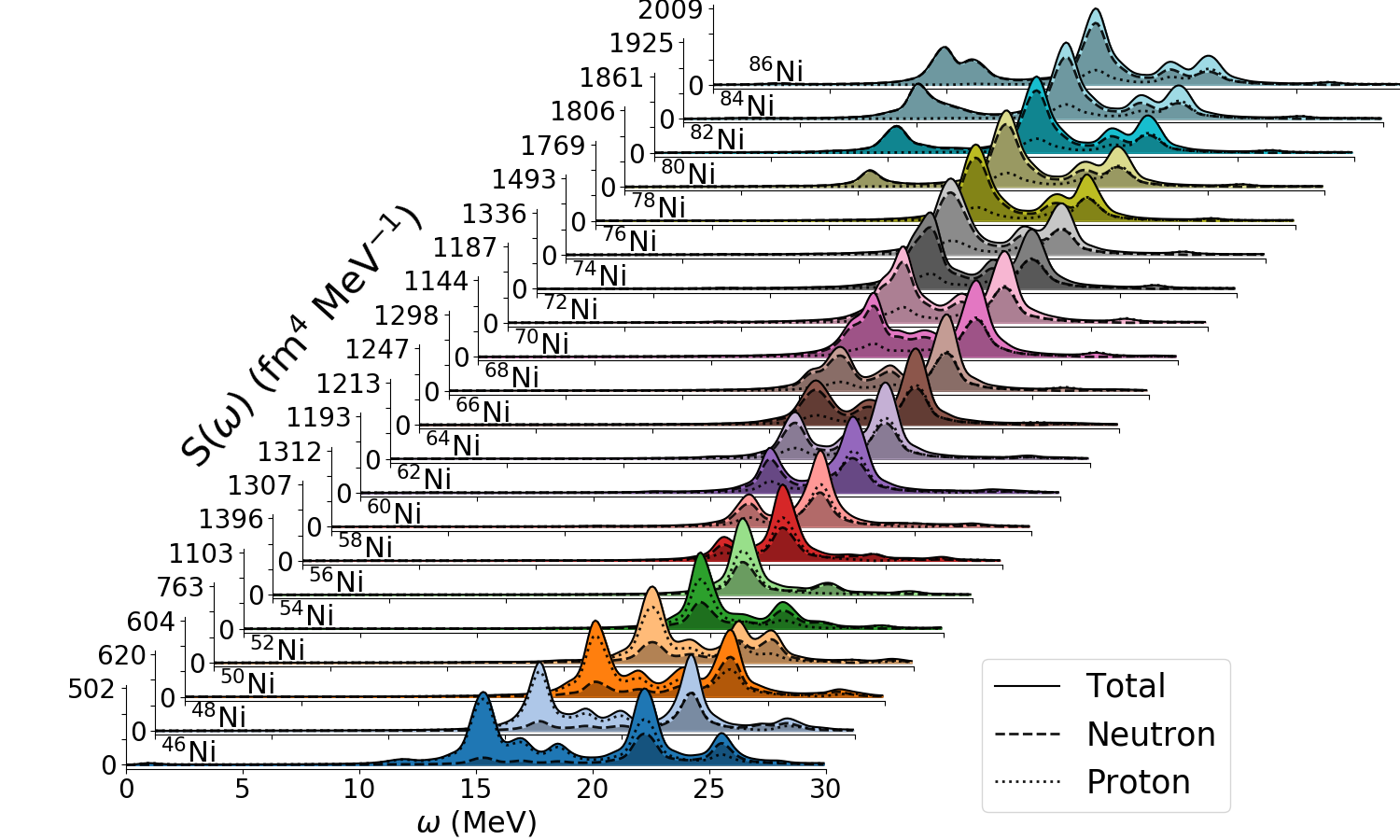}
%\par\end{centering}
\caption{Same as Fig.~\ref{fig:Ca_ISM_GS} but for Ni isotopes.}
\label{fig:Ni_ISM_GS}
\end{figure*}
%............................................................

\begin{figure}[!htb]
\centering
\includegraphics[width=1\linewidth]{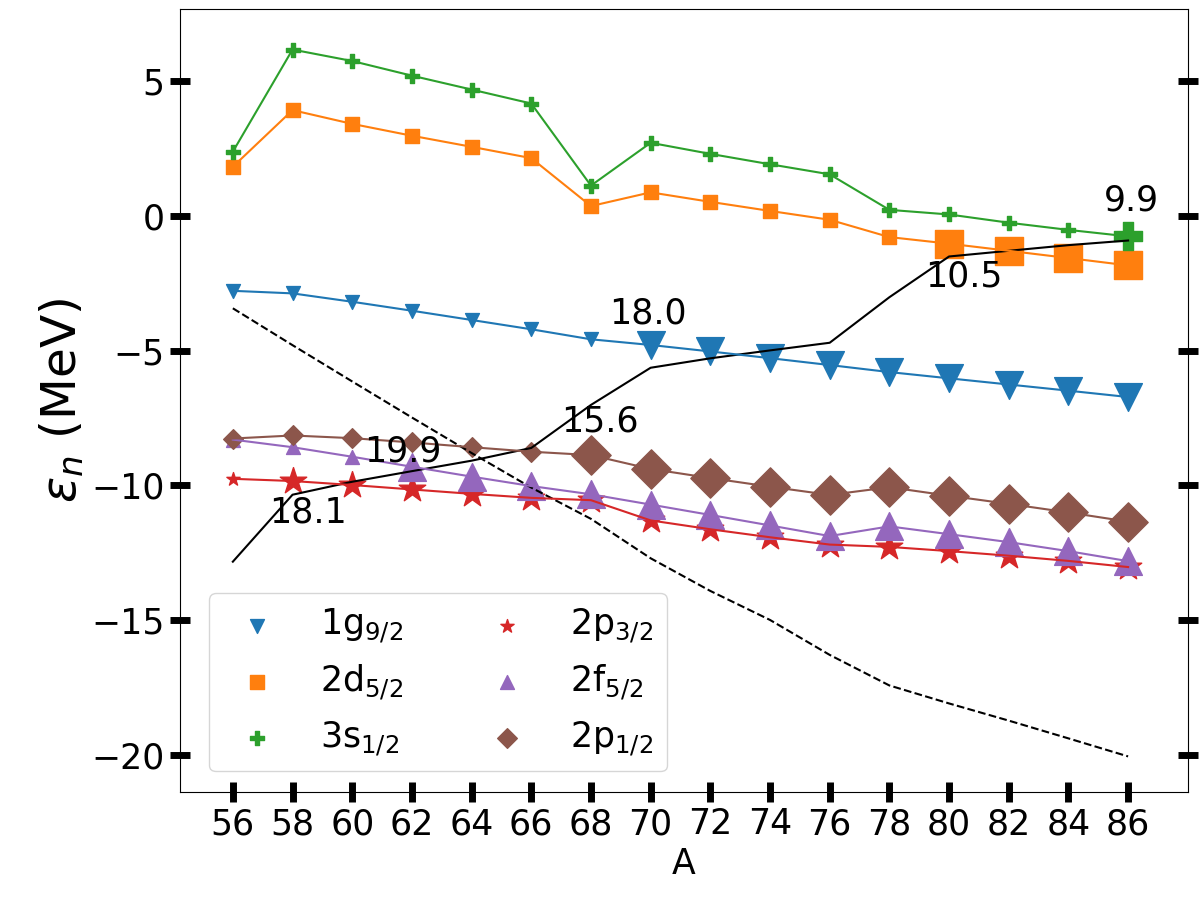}
%\par\end{centering}
\caption{Same as Fig.~\ref{fig:SP_Ca} but for Ni isotopes.}
\label{fig:Ni_appear}
\end{figure}

The previously discussed pattern of emergence of low-energy resonances, with increasing isospin asymmetry, can also be tested in the Nickel isotopic chain. Fig.~\ref{fig:Ni_ISM_GS} displays the monopole strength of $^{46-86}$Ni, obtained within the covariant QFAM. First focusing on the neutron-rich Ni isotopes, we expect new low-lying structures to emerge at $N=30$ and $N=52$, where orbitals that inaugurate new major shells start to be filled, i.e the $2p_{3/2}$ and the $2d_{5/2}$ sates, respectively. As illustrated in Fig.~\ref{fig:Ni_appear}, filling the $2p_{3/2}$ to $1g_{9/2}$ orbitals, yields resonances with energies around 18 MeV, while the occupation of the $2d_{5/2}$ and $3s_{1/2}$ states, gives rise to monopole excitations around 11 MeV.

%............................................................
%............................................................

\begin{figure}[!htb]
\centering
\includegraphics[width=1\linewidth]{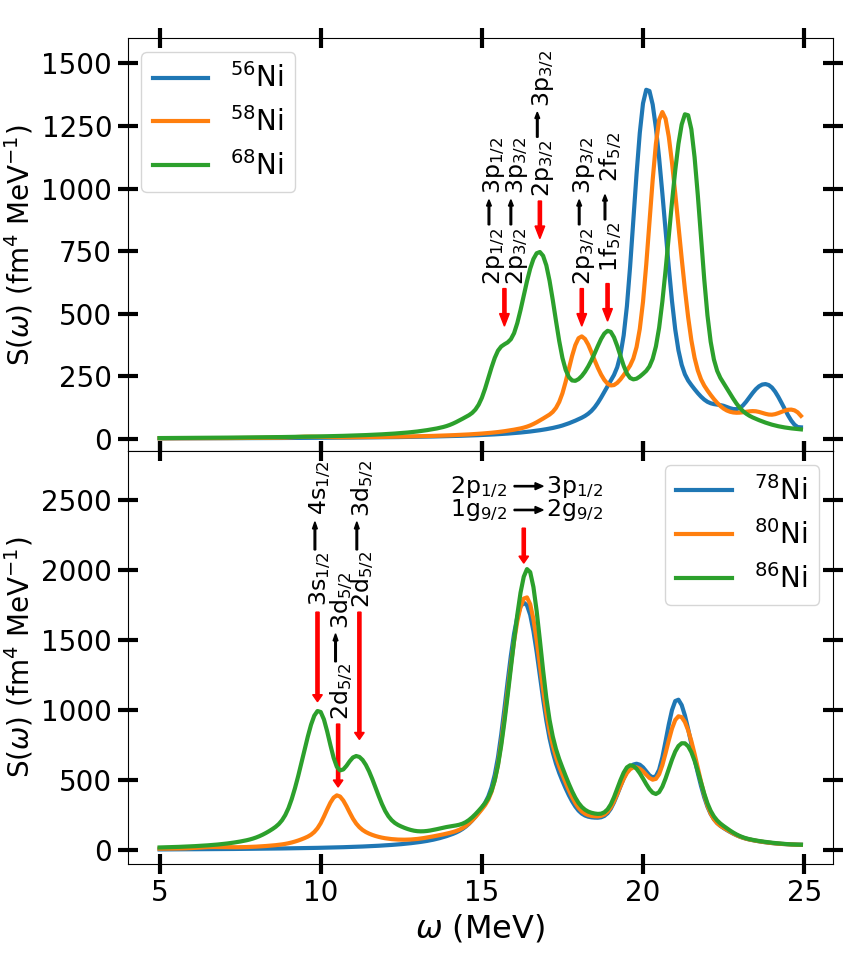}
%\par\end{centering}
\caption{Monopole strength of $^{56,58,68,79,80,86}$Ni, with the main 2qp contributions indicated for the low-energy modes.}
\label{fig:Ni_ISM_details}
\end{figure}

To better understand the evolution of the monopole strength in the neutron-rich $^{58,68,79,80,86}$Ni isotopes, Fig.~\ref{fig:Ni_ISM_details} relates each low-energy modes with its dominant 2qp contributions, while Tab.~\ref{tab:Ni_qp} further details the energy gap between the two single-particle states of the transition, as well as the weight of the 2qp configuration.
Hence, a new major shell, inaugurated by the $2p_{3/2}$ orbital, gets filled, going from $^{56}$Ni to $^{58}$Ni. Concomitantly, a new low-energy mode emerges at $\omega = 18.1$ MeV, dominated by the $2p_{3/2} \rightarrow 3p_{3/2}$ transition (42\%, see Tab.~\ref{tab:Ni_qp}). Likewise, in $^{62,68,70}$Ni, the occupation of the $2f5/2$, $2p1/2$ and $1g9/2$ orbitals comes with new resonances, located at $\omega = 19.9$ MeV, $\omega = 15.6$ MeV and $\omega = 18$ MeV, 
respectively. Their dominant 2qp contributions come from the $1f_{5/2} \rightarrow 2f_{5/2}$,  $2p_{3/2}\rightarrow 3p_{3/2}$ and $1g_{9/2} \rightarrow 2g_{9/2}$ transitions, respectively.
Adding more neutrons in the $2d_{5/2}$ and $3s_{1/2}$ orbitals, opens a new major shell, and thus new peaks appear in $^{80,86}$Ni at $\omega = 10.5$ MeV, and $\omega = 9.9$ MeV. Their dominant 2qp contributions come from the $2d_{5/2}\rightarrow 3d_{5/2}$ (32\%) in the former, and are more evenly distributed between the  $2d_{5/2}\rightarrow 3d_{5/2}$ (12\%) and $3s_{1/2}\rightarrow 4s_{1/2}$ (14\%), in the latter.

%............................................................
\begin{table}
%\begin{tabular}{ |p{1cm} | p{1cm} | p{2.5cm} | p{1.5cm} | p{1.5cm}|  }
\begin{tabular}{ |P{12mm}|P{12mm}|P{22mm}|P{15mm}|P{15mm}|  }
 \hline
 Nucleus & $\omega$ (MeV) & Transition $\alpha \rightarrow \beta$ & $E_\beta + E_\alpha$ (MeV) & $\left|X_{\alpha \beta}\right|^{2}-\left|Y_{\alpha \beta}\right|^{2}$  \\
 \hline \hline  
 
 $^{58}$Ni & 18.1 & $2p_{3/2} \rightarrow 3p_{3/2}$ & 16.0 & 0.42 \\
 & & $2p_{1/2} \rightarrow 3p_{1/2}$ & 15.4 & 0.02 \\
 \hline
  
 $^{68}$Ni & 15.6 & $2p_{1/2} \rightarrow 3p_{1/2}$ & 11.8 & 0.33 \\
 & & $2p_{3/2} \rightarrow 3p_{3/2}$ & 10.0 & 0.13 \\
 \hline
 
 $^{70}$Ni & 18.0 & $2p_{3/2} \rightarrow 3p_{3/2}$ & 7.0 & 0.22 \\
 & & $1g_{9/2} \rightarrow 2g_{9/2}$ & 15.9 & 0.19 \\
 
 \hline
 
 $^{80}$Ni & 10.5 & $2d_{5/2} \rightarrow 3d_{5/2}$ & 9.5 & 0.32 \\
 & & $3s_{1/2} \rightarrow 4s_{1/2}$ & 9.0  & 0.06 \\
 & & $2d_{3/2} \rightarrow 3d_{3/2}$ & 9.1  & 0.04 \\

 \hline
 
 $^{86}$Ni & 9.9 & $3s_{1/2} \rightarrow 4s_{1/2}$ & 8.4 & 0.14 \\
 & & $2d_{3/2} \rightarrow 3d_{3/2}$ & 8.0 & 0.12 \\
 \hline

\end{tabular}
 \caption{Same as Tab.~\ref{tab:Ca_qp} for Ni isotopes.}
 \label{tab:Ni_qp}
\end{table}

From a general point of view, the evolution of the monopole strength, in both Ca and Ni isotopic chains, are quite similar, mainly driven by the single-particle spectrum features. We have checked that these similarities are also present at the level of the transition densities.
Interestingly, the low-energy part of the $^{68}$Ni monopole strength is in agreement with the experimental results reported in Ref.~\cite{van14}, where a peak around 15 MeV is measured.    
Also, the structure of the strength obtained in $^{68-78}$Ni isotopes is in agreement with the calculations reported in Ref.~\cite{ham14}, where the coupling to the continuum was considered. Therefore, this last effect does not impact the qualitative features of the monopole strength. However, it strongly impacts the width of the low-energy resonances, and enhances their collectivity. 
Finally, the proton counterpart of low-energy modes, can be observed in the  neutron-deficient $^{46-56}$Ni (see Fig.~\ref{fig:Ni_ISM_GS}), with the same mechanism of emergence. Decreasing the neutron number drives the proton Fermi energy towards the continuum, decorrelating protons that used to be bound, into a core in the case of the neutron-rich isotopes. As a result, the neutron contribution to the monopole strength between 15 and 20 MeV decreases, and new purely proton modes emerge, e.g. at $\omega = 15.5$ MeV  and $\omega =17.3$ MeV in $^{46}$Ni.

%............................................................
%............................................................

\subsection{Soft modes contribution to the total strength}

\begin{figure}[!htb]
\centering
\includegraphics[width=1\linewidth]{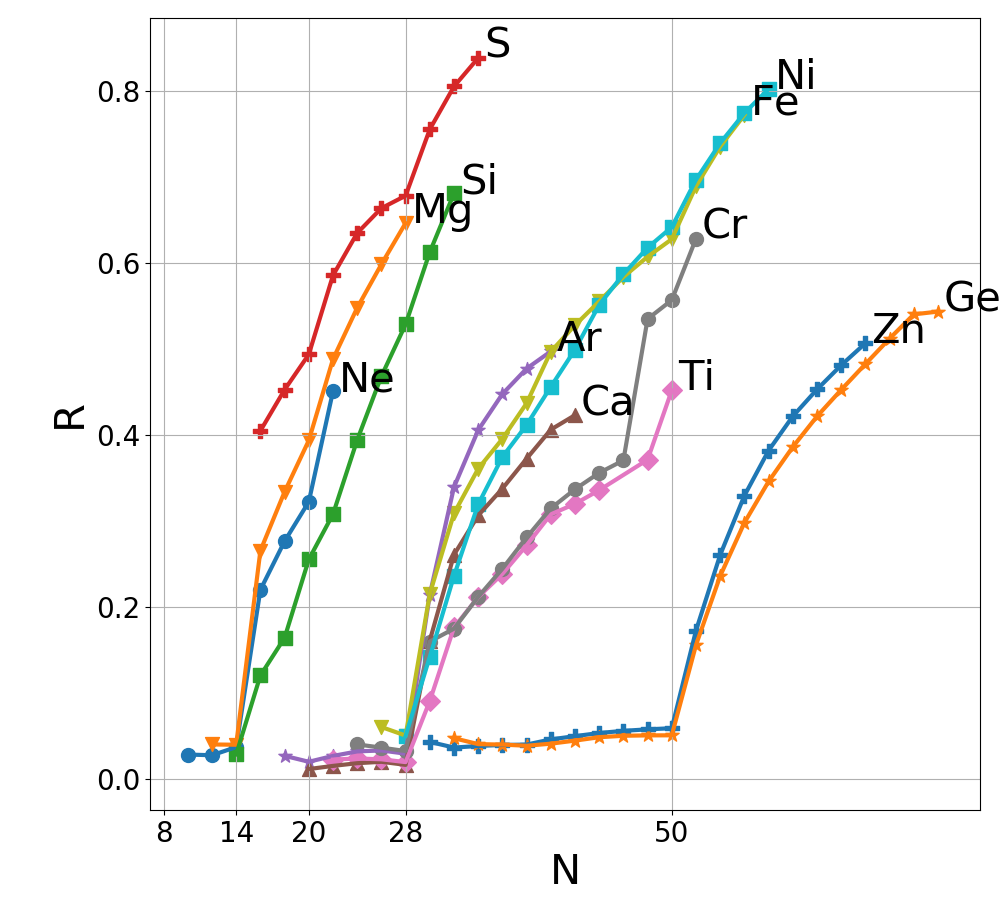}
%\par\end{centering}
\caption{Evolution of the ratio~\eqref{eq:ratio}, with the neutron number $N$, for various isotopic chains. The integration over soft neutron modes is performed up to $\omega_\text{sm} \sim 14-18$ MeV, depending on the position of the GR for each nuclei.}
\label{fig:ratio_ISM}
\end{figure}

A possible way for quantifying the evolution of soft modes with isospin asymmetry, is to evaluate their global contribution to the total strength.
On this purpose, Fig.~\ref{fig:ratio_ISM} displays the evolution of the ratio of integrated strengths of the soft modes to the total one:
\begin{equation}
    R\equiv \frac {\int _0 ^{\omega_\text{sm}}d\omega S(\omega) } {\int _0 ^{\omega_\text{tot}}d\omega S(\omega)},
\label{eq:ratio}    
\end{equation} 
where $\omega_\text{sm}$ stands for the energy of the last soft mode, and $\omega_\text{tot} = 30$ MeV. This ratio is displayed for various isotopic chains, ranging from Ne to Ge. It should
be noted that deformed nuclei have been constrained to a spherical shape, since it is relevant 
to keep the study of the impact of deformation, for the next section. When there are multiple soft modes,separated by few MeV, the sum of the soft modes up to the last one is considered. 

%............................................................
%............................................................

The general trend corresponds to a systematic increase of the proportion of the soft mode in the total monopole strength with neutron excess. This ratio can reach close to 80\% of the total strength for Nickel, Iron or Sulfur very neutron-rich nuclei.
Here again, the relation between the opening of a major shell, and the emergence of soft neutron modes, is visible, i.e. adding neutrons on top of configurations with  $N=14,28,50$ yields a sudden increase of the contribution of soft modes. On the other hand, adding neutrons on top of the $N=20$ shell closure has little impact, although an inflexion in the evolution of the soft modes proportion can be observed in Sulfur and Neon isotopic chain (Fig.\ref{fig:ratio_ISM}). In the case of Argon and Calcium isotopes, the mismatch between the proton and neutron Fermi energies, is not large enough, to trigger a low-energy resonance at $N=22$. 

Another feature, is that substructures stemming from the opening of subshells, have negligible effects in the evolution of this ratio: when the occupation of an orbital opens a new possible monopole transition, the studied ratio is not impacted. However, one exception occurs with the $2s_{1/2}$ orbital: its filling gives rise to the first appearance of neutron soft mode in Neon, Magnesium, Silicone and Sulfur at $N=16$. This may be related to the fact that $N=14$ is a strong subshell closure~\cite{bec06}.

%............................................................
\begin{figure*}[!htb]
\centering
\includegraphics[width=1\linewidth]{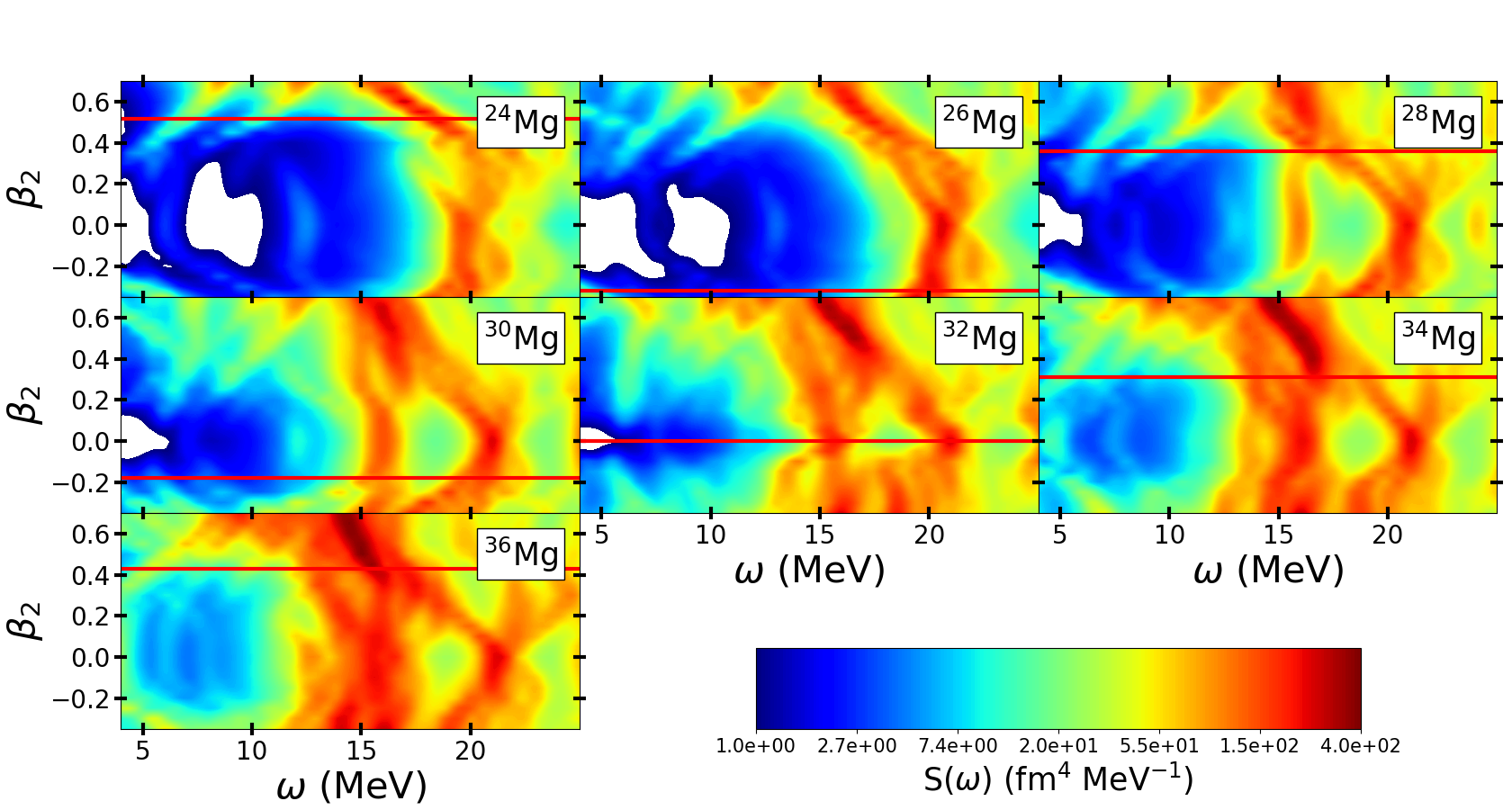}
%\par\end{centering}
\caption{Evolution of the monopole strength distribution, in Mg isotopes, with the deformation of the reference RHB state, on top of which the QRPA response is built. The red lines show the ground state deformation, for each nuclei at the RHB level. }
\label{fig:MAP_Mg_ISM}
\end{figure*}
%............................................................

%
%===================================================================================================================
\section{\label{sec:def}  Evolution of low-energy isoscalar monopole modes with deformation}
%===================================================================================================================
%
We now wish to investigate the impact of deformation, on the structure of the low-energy part of the monopole strength. The splitting of these modes, with the onset of deformation, is a known feature \cite{pen09}. The role played by deformation, in the appearance of so-called cluster modes at very low energy, was also studied in details in $N=Z$ nuclei~\cite{mer21}. The aim of this section, is to extend the analysis to neutron-rich systems.  In order to assess how nuclear deformation affects the structure of the monopole strength, we performed QFAM calculations on top of RHB reference states, constrained to different deformations, parametrized by the axial quadrupole parameter $\beta_2\in[-0.2,0.8]$. This allows to monitor the various monopole resonances, as the deformation of the system is changed, both for soft and cluster modes, each of them addressed separately in the next two subsections.

\subsection{Evolution of soft modes with deformation and neutron excess}

The evolution of soft modes with deformation, and isospin asymmetry, involves generic patterns, which will be illustrated in the specific case of Mg isotopes. 
We start by examining the evolution of the monopole strength in Mg isotopes, as the constrained axial quadrupole deformation of the RHB reference state varies from $\beta_2 = -0.2$ to $\beta_2 = +0.8$ (Fig.~\ref{fig:MAP_Mg_ISM}). The evolution of the monopole strength with neutron excess, in spherically-constrained Mg isotopes ($\beta_2=0$ slice in each panel of Fig.~\ref{fig:MAP_Mg_ISM}),
is in agreement with the previous discussion on neutron low-energy modes emergence. Namely, a low-energy mode appears in $^{28}$Mg, related to the filling of the $2s_{1/2}$ orbitals.
Then, the width of the strength increases in $^{30}$Mg, due to an additional excitation from $1d_{3/2}$ to $2d_{3/2}$. In $^{34,36}$Mg, a new soft mode at $\omega \approx 13.5$ MeV emerges, stemming from the filling  of the $1f_{7/2}$ orbital.

\begin{figure*}[!htb]
\centering
\includegraphics[width=1\linewidth]{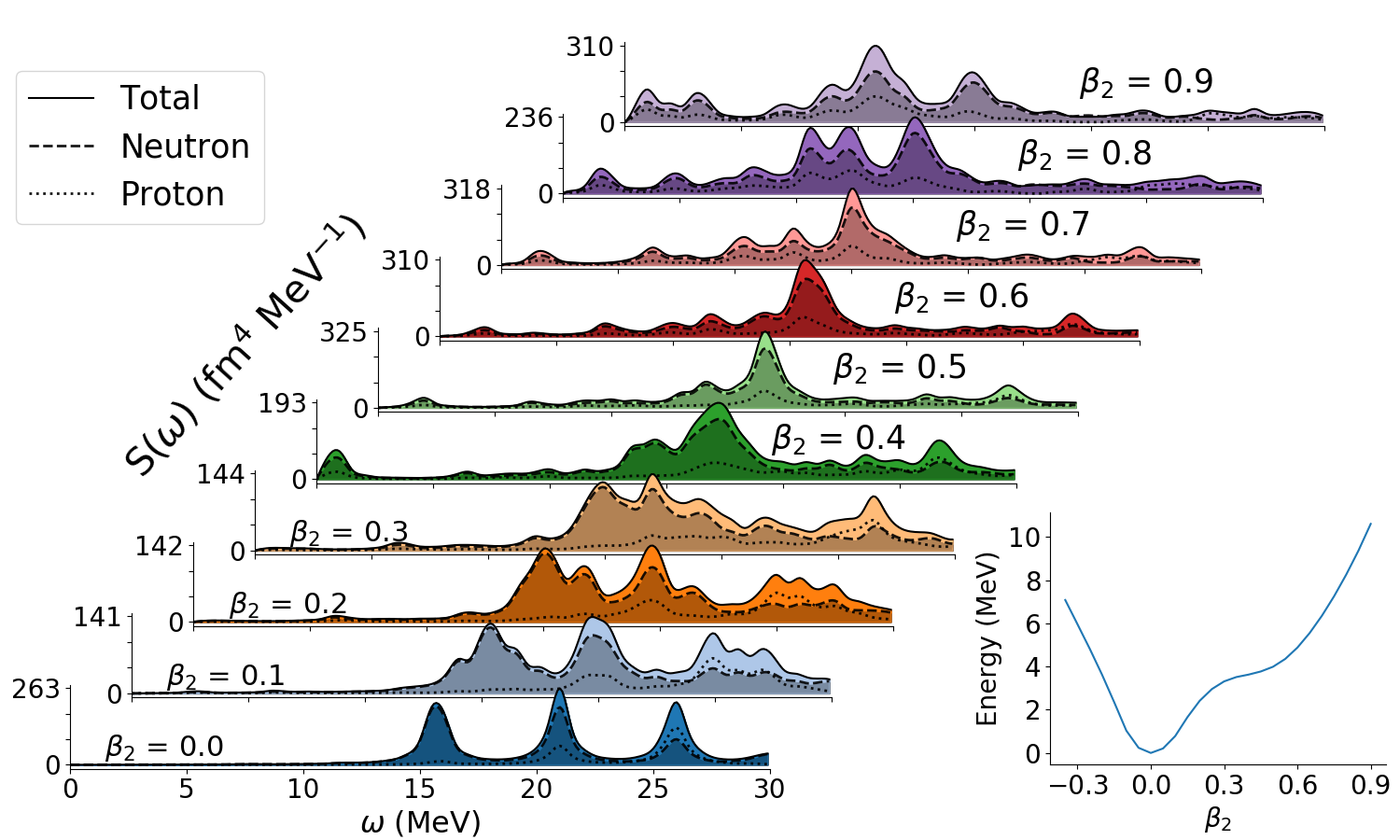}
%\par\end{centering}
\caption{Left : evolution of $^{32}$Mg monopole strength distribution with the deformation of the reference RHB state, split into total (solid line), neutron (dashed line) and proton (dotted line) contributions. Right :  total energy curve of $^{32}$Mg, with respect to the axial quadrupole deformation parameter $\beta _2$.} 
\label{fig:Mg32_1d}
\end{figure*}

The impact of deformation, in the low-energy part of the monopole strength, can be studied by tracking the soft modes as the deformation switches on, and increases. Focusing on $^{32}$Mg (see the corresponding panel in Fig.~\ref{fig:MAP_Mg_ISM} as well as in Fig.~\ref{fig:Mg32_1d}), the onset of deformation causes a splitting of the soft modes, similarly to what is known for the GMR. For instance, the resonance found at $\omega=15.5$ MeV, in the spherical $^{32}$Mg, splits into four components in the deformed $\beta_2=0.1$ case, located at $\omega =$14.0, 15.4, 16.5 and 17.5 MeV, respectively. A fifth component is visible at $\omega \approx 11.8$ MeV, the latter acquiring more strength as the deformation increases.

%............................................................
%...........................................................

\begin{figure}[!htb]
\centering
\includegraphics[width=1\linewidth]{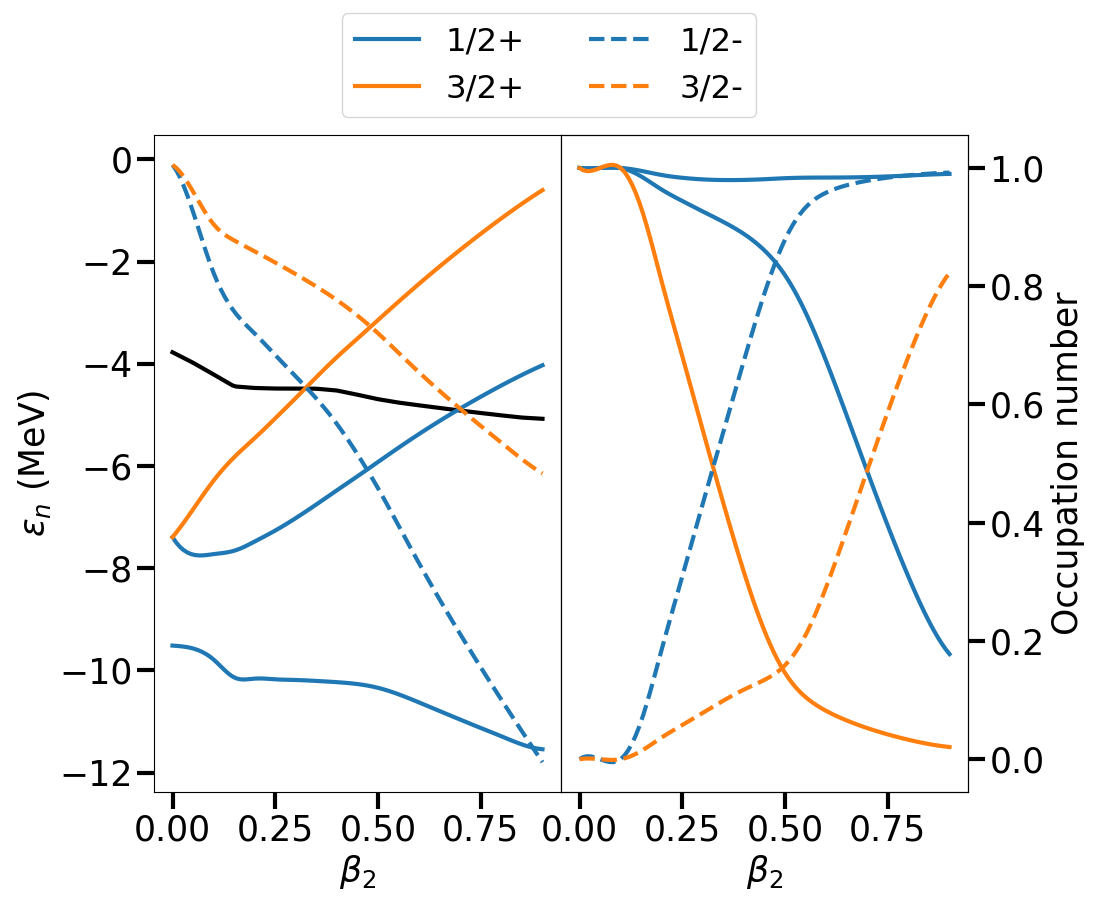}
%\par\end{centering}
\caption{Subset of $^{32}$Mg neutron canonical single-particle energies (left), and their occupation number (right), plotted against the axial quadrupole deformation parameter $\beta_2$. The black line shows the neutron Fermi level.}
\label{fig:Mg32_e_o}
\end{figure}

Because of the non-collective nature of these soft modes, one can draw a correspondence between the splitting in the strength, and the splitting induced by the axial deformation at the level of the canonical single-particle spectrum (Fig.~\ref{fig:Mg32_e_o}). For instance, the dominant 2qp contributions of the 15.7 MeV monopole mode of $^{32}$Mg, at $\beta _2 =0$ (see Fig.~\ref{fig:Mg32_1d}), are $2s_{1/2}\rightarrow 3s_{1/2}$ and $1d_{3/2}\rightarrow 2d_{3/2}$.
With the onset of deformation, the spherical $1d_{3/2}$ orbital splits into non-degenerate $\Omega^\pi = 1/2^+$ and $\Omega^\pi = 3/2^+$ states, where $\Omega$ stands for the projection of the total angular momentum $J$ on the symmetry axis (chosen to be the (Oz) axis) and $\pi$ the parity of the state. With the breaking of the rotational symmetry, these states do not belong to an irreducible representation of the $SU(2)$ group labeled by the eigenvalues of $J^2$, but rather mix the $m=\Omega$-component of positive parity spherical orbitals (provided reflection symmetry remains unbroken): the  $\Omega^\pi = 1/2^+$ axially-symmetric state results from the mixing of the $m=1/2$ component of the $s$, $d$, $g$, etc. orbitals.       

In $^{32}$Mg, as the $1d_{3/2}$ orbital splits into a $\Omega^\pi = 1/2^+$ and a $\Omega^\pi = 3/2^+$ states,  new transitions take place between the occupied $1/2^+$ (resp. $3/2^+$) and all the other unoccupied $1/2^+$ (resp. $3/2^+$) begotten by the $SU(2)$ symmetry breaking. Therefore, much more transitions are available, as compared to the spherical case, causing an enhancement of the collectivity of the low-energy resonances. For instance, in the case of $^{32}$Mg at $\beta _2 =0.1$, between $\omega=11$ MeV and $\omega=16$ MeV, almost 10 transitions are involved, and carry a non-negligible part of the strength (more than 5\% of the total strength each).
In appears that what is called a monopole mode, in such a deformed framework, should be understood as a mixing between monopole, quadrupole, etc. transitions.

%............................................................
%...........................................................

In the case of larger deformation, Fig.~\ref{fig:Mg32_1d} shows that the GMR is shifted to lower energy, as expected, and starts to merge with the soft modes. However, the distinction between soft modes, and GMR, can be made by looking to the proton contribution to the total strength (dotted lines on Fig.~\ref{fig:Mg32_1d}). When the latter is non-zero, there is a coherent excitation of both protons and neutrons, corresponding to the GMR. Based on this criteria, we deduce that the soft modes vanish for $\beta _2 >0.7$, due to the spreading of the GMR (Fig.\ref{fig:Mg32_1d}).

\begin{figure}[!htb]
\centering
\includegraphics[width=1\linewidth]{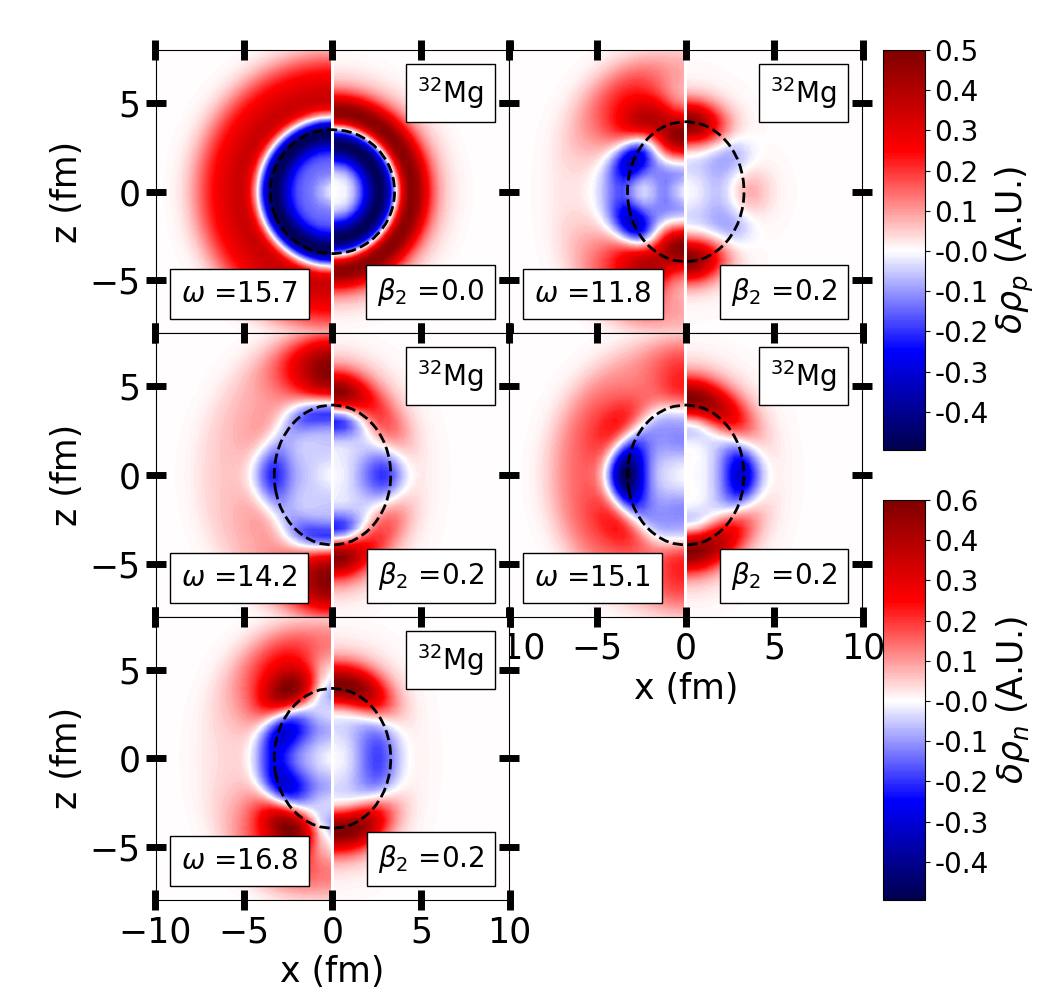}
%\par\end{centering}
\caption{Neutron (left part of each panels) and proton (right part of each panels) transition densities, associated to resonances with excitation energy $\omega$ (in MeV) of $^{32}$Mg constrained at deformation $\beta_2$. The black dashed lines show the contour of the unperturbed total density (solution of the static, constrained RHB equation) around which the nucleus vibrates.}
\label{fig:Mg32_dens}
\end{figure}

Transition densities are shown in Fig.~\ref{fig:Mg32_dens},
for the mode located at $\omega = 15.7$ MeV in the spherical $^{32}$Mg, as well as for the modes at $\omega = 11.8,14.2,15.1,16.8$ MeV in $^{32}$Mg, constrained to $\beta _2 = 0.2$. Deformation generates localisation 
on the transition densities, as shown by the proton one in the core of the nucleus, and the neutron one both in the core and in the surface of $^{32}$Mg.

%............................................................
%...........................................................

The spatial properties of the transition densities, shown in Fig.~\ref{fig:Mg32_dens}, can be further analyzed in terms of the 2qp contributions to the corresponding modes, and of the shape of the canonical orbitals, involved in the corresponding monopole transitions.
%............................................................
\begin{figure*}[!htb]
\centering
\includegraphics[width=1\linewidth]{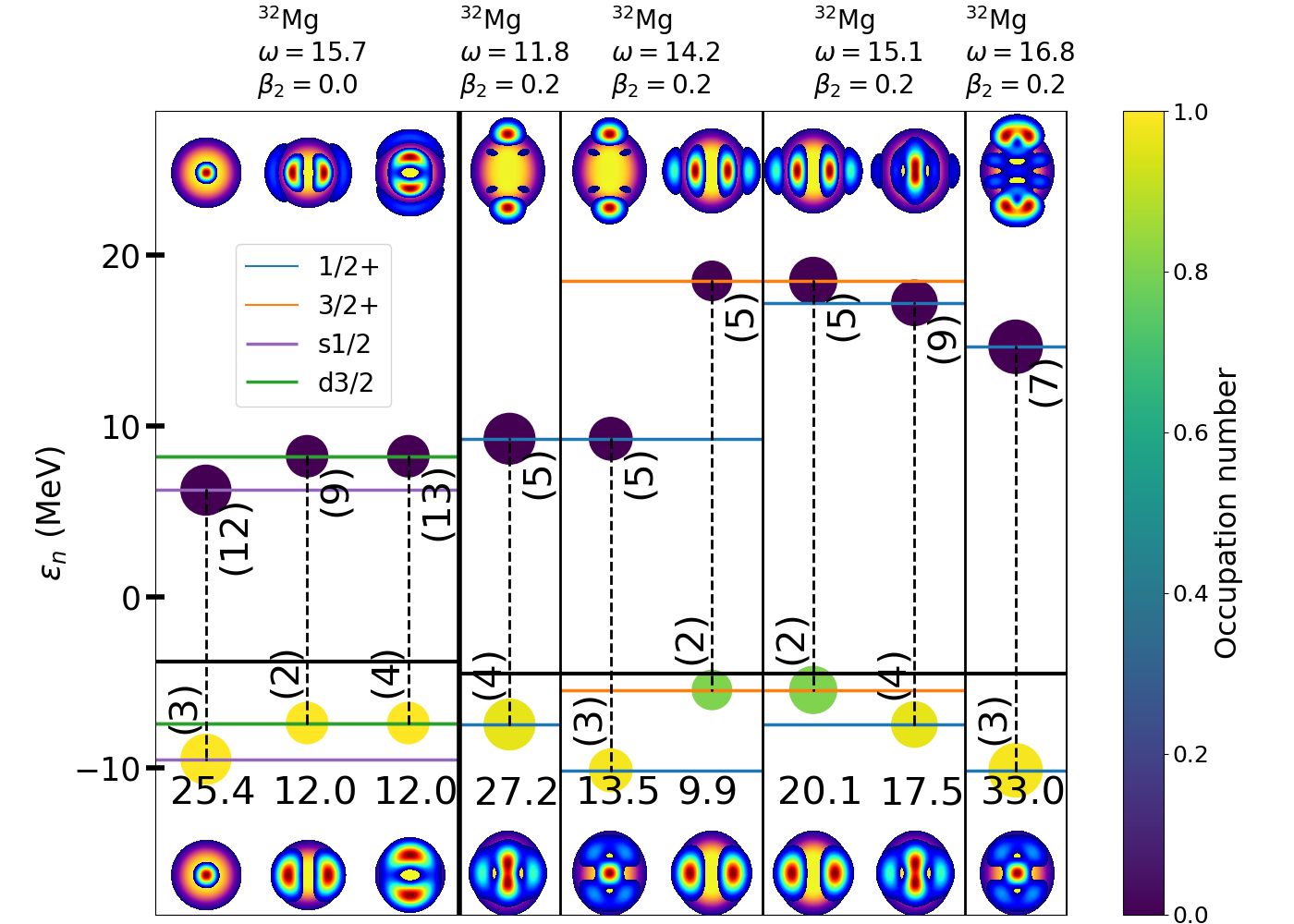}
%\par\end{centering}
\caption{Canonical neutron orbitals, involved in the dominant monopole transitions contributing to the modes found at excitation energy $\omega$ (in MeV) in $^{32}$Mg, with axial quadrupole deformation $\beta_2$. The percentage to which the transition contribute to the monopole strength, is indicated by the number and the surface area of the circle on top of the level, with a color giving the occupation number of the orbital. The number in parenthesis shows principal quantum number of each levels. Partial densities of the orbitals, involved in the monopole transitions, are plotted on top of the total nucleon density.}
\label{fig:Mg32_cano}
\end{figure*}
%...........................................................
The first column of Fig.\ref{fig:Mg32_cano} corresponds to the monopole mode at $\omega = 15.7$ MeV, in the spherical $^{32}$Mg. It displays the canonical states participating to the dominant 2qp configuration, i.e. 
the 2 and 3$s1/2$ orbitals, as well as the 1 and 2$d3/2$ orbitals. The corresponding configurations contribute to about 50\% to the monopole strength. 
The second column of Fig.\ref{fig:Mg32_cano} corresponds to the excitation at $\omega = 11.8$ MeV in $^{32}$Mg, constrained at $\beta_2=0.2$. The main 2qp contribution to this mode involves the 1/2+(4) state, which coincides with the $m=1/2$ component of the $1d3/2$ in the $\beta_2=0$ limit. Here, only some features of the transition density can be related to the 2qp configuration, namely the localisation of the neutron transition density on the radial axis, visible on the 4th 1/2+ orbital. It should be noted that  one cannot expect a full correspondence between the main 2qp configurations and the transition densities, since the contribution of the former is about 30\%.

The third and fourth columns of Fig.\ref{fig:Mg32_cano}, correspond to the $\omega = 14.2$ MeV and $\omega = 15.1$ MeV monopole modes, found in $^{32}$Mg constrained at $\beta_2$=0.2. Both modes are slightly more collective than the ones located at $\omega = 11.8$ MeV and $\omega = 16.8$ MeV. The major 2qp contribution to the resonance located at 14.2 MeV comes from the 1/2+(3) $\rightarrow$ 1/2+(5) transition, where the 1/2+(3) state coincides with the $2s1/2$ orbital, in the $\beta_2=0$ limit. In this case again, the shape of the partial densities associated to this sole transition allows to understand the main spatial properties of the corresponding transition density in Fig.~\ref{fig:Mg32_dens}, and to interpret this mode as a cluster vibration. 
On the other hand, the excitation found at $\omega =15.1$ MeV is dominated by the 3/2+(2) $\rightarrow$ 3/2+(5) transition, where the 3/2+(2) state coincides with the $m=3/2$ component of the $1d{3/2}$ orbital in the $\beta _2 = 0$ limit, together with a contribution coming from the 1/2+(4) $\rightarrow$ 1/2+(9) transition. In that case also, one can trace back the spatial properties of the corresponding neutron transition density (Fig.~\ref{fig:Mg32_dens}), namely a pronounced contribution on the horizontal axis, to the shape of the canonical partial densities. Finally, the last column of Fig.\ref{fig:Mg32_cano} shows the mode at $\omega$=16.8 MeV, where the state matching the 1/2 component of the 2s$_{1/2}$, in the $\beta _2 \rightarrow 0$ limit, is involved. 
The main corresponding 2qp contribution, comes from the 1/2+(3)$\rightarrow$ 1/2+(7) transition, with again a shape of the transition density, that can be understood by looking at the canonical partial densities.

\subsection{Evolution of cluster excitation with deformation and neutron excess}

Cluster vibrations, described as coherent excitation of neutrons and protons localized in clusters, are expected to occur in $N=Z$ nuclei at low energies --- typically between 5 and 10 MeV --- and large deformations~\cite{mer21}. In this subsection, we investigate the impact of neutron excess on these modes. 

For all the Mg isotopes considered in Fig.\ref{fig:MAP_Mg_ISM}, there is a systematic occurrence of low-energy modes, starting from $\beta_2\sim 0.4$. The analysis of these modes, in terms of 2qp configurations, and the computation of the time-dependent density, allows to tag them as cluster oscillations. 
Let us focus again on one typical example, from $^{32}$Mg.  
In Fig.\ref{fig:Mg32_1d}, a new structure in the monopole strength starts to develop around $\omega = 6$ MeV, from $\beta_2\sim 0.2$ and increases with deformation. The emergence of this mode can be traced back to the shell structure of $^{32}$Mg in Fig. \ref{fig:Mg32_e_o}: 
the 1/2-(3) and 3/2-(2) states, responsible for localizing neutrons in clusters along the symmetry axis, quickly falls towards the Fermi energy, as the quadrupole deformation increases. They even become fully occupied from $\beta_2\sim 0.4$ for the former, and $\beta_2 \sim 0.7$ for the latter (see also the corresponding occupation numbers, plotted against $\beta_2$ in the right panel of Fig.\ref{fig:Mg32_1d}). 
As for proton orbitals, the dominant transitions contributing to the mode at $\omega\sim 6$ MeV are the same for all Mg isotopes, namely 1/2+(2)$\rightarrow$1/2+(3), 1/2-(2)$\rightarrow$ 1/2-(3) and 3/2+(1)$\rightarrow$3/2+(3). Again, these levels are the ones localizing the protons in clusterized structures along the symmetry axis.

\begin{figure}[!htb]
\centering
\includegraphics[width=1\linewidth]{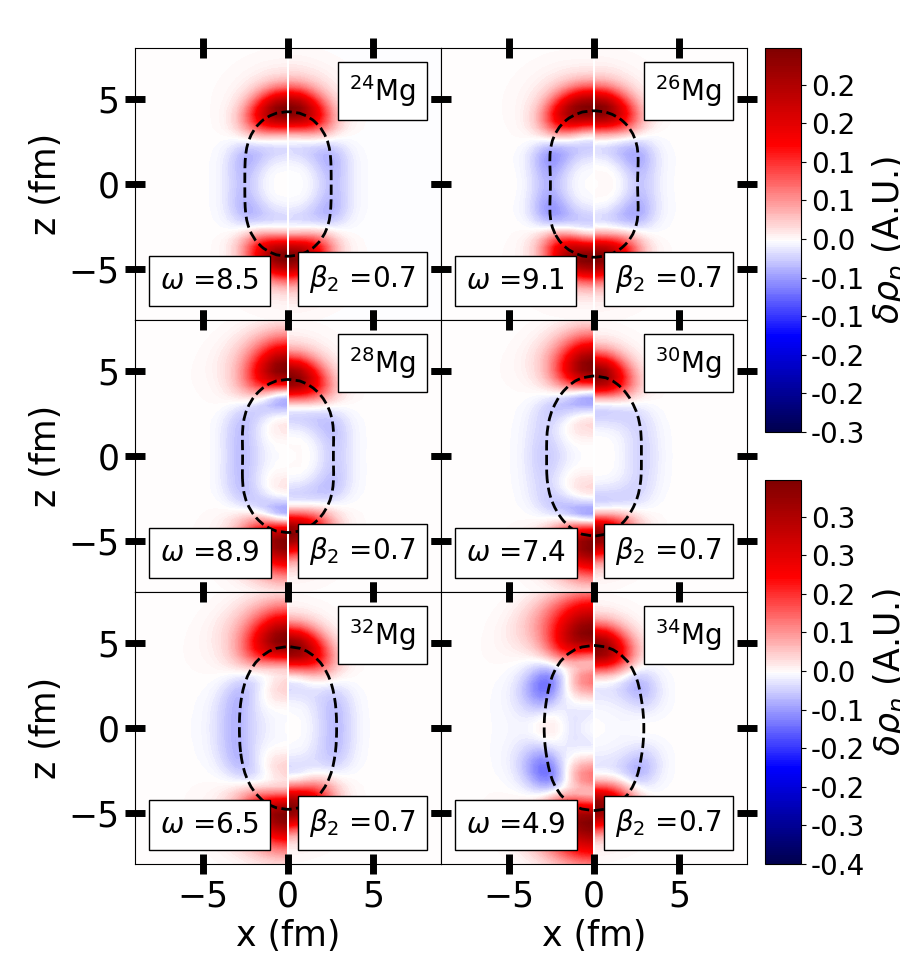}
\caption{(color online).Imaginary part of the transition densities for some Magnesium isotopes. Excitation energies are given in MeV. Deformation is constrained to take the value of $\beta _2=0.7$ for all nuclei. On the left of each plot, is shown the neutron density, and on the right the proton one. The dashed black lines represent the RMS radius for matter ground state.}
\label{fig:Mg_clusters}
\end{figure}

In order to understand the interplay between deformation and neutron excess in cluster modes,
the transition densities of these modes in $^{24-32}$Mg isotopes is displayed in Fig.\ref{fig:Mg_clusters}.
The $^{12}$C+$^{12}$C structure in the excited state, is seen on the transition densities of $^{24}$Mg, where the neutron and proton oscillate in phase, in the clusters location. When the neutron number increases, the neutron and proton transition densities gets slightly shifted with respect to each other, but the main effect occurs closer to
the center of the nucleus: additional peaks appear in the neutron transition density. This implies more complex
vibrations than a mere oscillation of the clusters, namely small additional contributions to the excitation, mainly at an average distance between the center and the surface of the nucleus, on the symmetry axis. However, it should be noted that the $^{12}$C+$^{12}$C oscillation is still present in all the considered isotopes.

%............................................................
%...........................................................

%
%===================================================================================================================
\section{\label{sec:pairing}  Evolution of low-energy isoscalar monopole modes with pairing}
%===================================================================================================================
%

As mentioned in the case of $^{62}$Ca, another kind of low-energy mode appears below 5 MeV, driven by pairing correlations. Such excitations have been discussed both 
theoretically and experimentally, see e.g. \cite{bes66,bro77,vit80}. Pairing vibration modes refer to coherent excitations involving particle-particle, particle-hole and hole-hole contributions. Their properties were analysed within the QRPA framework both for spherical~\cite{kha04,mat04,shi11} and deformed~\cite{yos06} nuclei. 
In the present work, we investigate the  possible interplay between pairing and cluster modes.

%===================================================================================================================
\subsection{Spherical case}
%==========================================================

How pairing correlations influence the monopole strength distribution can be studied along the same lines as in the deformed case, i.e. by constraining the amount of pairing correlation captured by the reference RHB state and then monitoring the monopole strength, as pairing correlations increase. Following Ref.~\cite{dug20}, the constraint on pairing is implemented by varying the strength of the pairing interaction, from $0$ to twice of its normal value, both at the RHB and QFAM levels.
We define the total pairing energy per nucleon ($E_\text{pair/A}$) as an acceptable order parameter for the normal to superfluid phase transition and plot our result against the latter. 

We first focus on the evolution of the ISM strength distribution with pairing correlation at zero deformation. As soon as the pairing energy reaches a threshold value, the last occupied orbitals will start to deplete, enabling low-energy pair excitations.

\begin{figure}[!htb]
\centering
\includegraphics[width=1\linewidth]{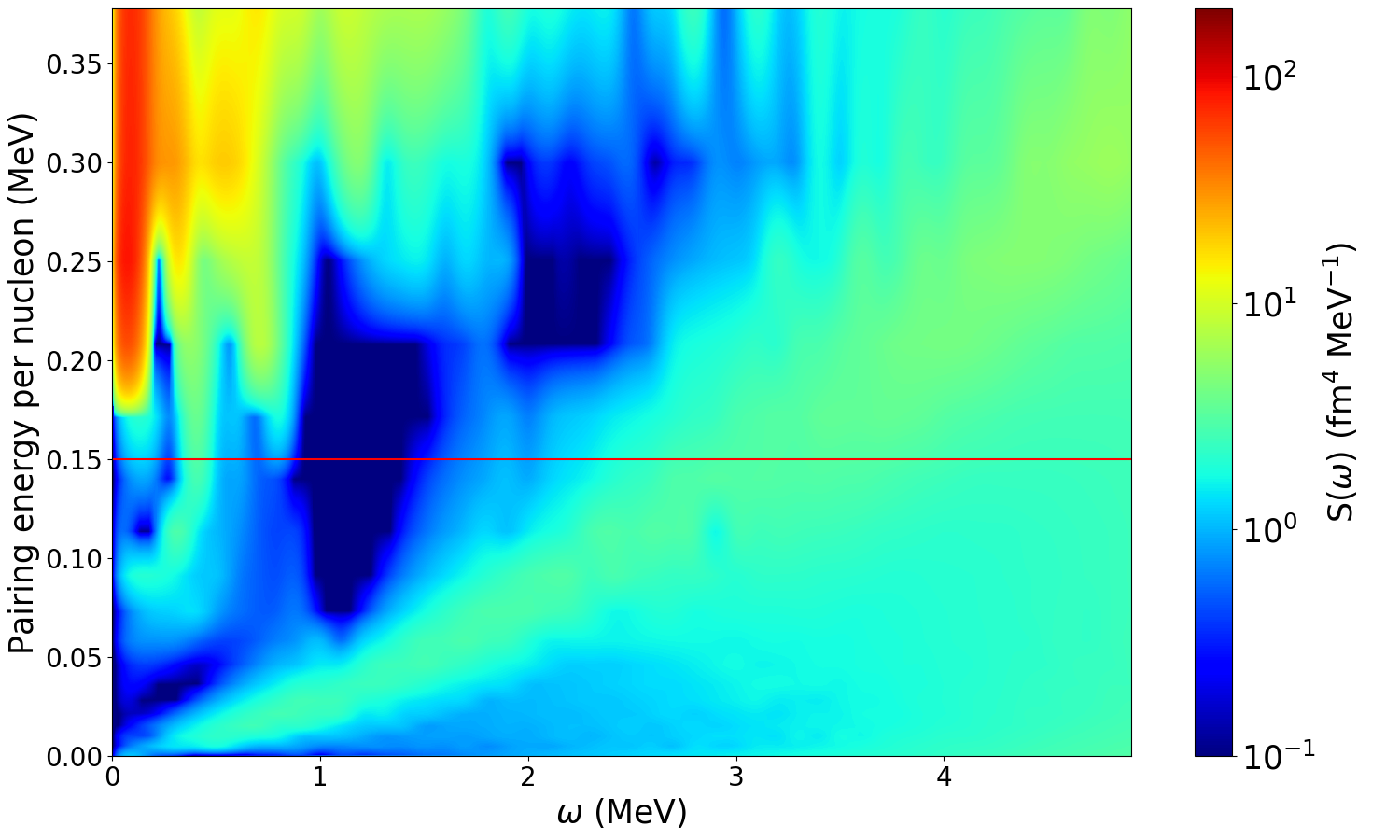}
\caption{(color online). Evolution of the monopole strength function of $^{54}$Ca with the pairing energy of the RHB level. The red horizontal line shows the pairing energy of the ground state $^{54}$Ca.}
\label{fig:Ca_pairing}
\end{figure}

This typical behavior is illustrated in the case of $^{54}$Ca. Fig.\ref{fig:Ca_pairing} shows the evolution of the ISM strength function with the pairing energy per nucleon associated to the RHB reference state (the total pairing energy in the RHB ground state being 7.6 MeV). With the increase of pairing correlations, a new mode appears below 1 MeV, stemming from the fact that the $2p_{1/2}$ orbital becomes partially occupied. This mode already appears in $^{54}$Ca ISM strength built on top the of RHB ground state, however with a negligible contribution: the maximum strength of $^{54}$Ca, corresponding to the GMR, is $\sim$600 fm$^4$ MeV$^{-1}$, while this low-energy mode has a peak of only $\sim$5 fm$^4$ MeV$^{-1}$. For large pairing correlations, this excitation carries more strength, up to  $\sim$100 fm$^4$ MeV$^{-1}$ within the range considered for the constrained calculations.

%............................................................
%...........................................................

This typical behavior is quite general and has been checked for several nuclei, both in the Calcium and Nickel isotopic chains. The contribution of pairing-type excitations to the total ISM strength remains very low, but increases if the pairing is constrained to larger values. 
The typical excitation energy range depends on the pairing energy, since this resonance can be considered as a pure pairing mode: the energy of the excitation is then expected to be of the order of $\sim 2\Delta$, where $\Delta$ stands for the pairing gap, explaining why the excitation energy of the pairing mode increases with the pairing energy itself.
A typical transition density is plotted on Fig.\ref{fig:Ca_td} in the bottom right corner for $^{62}$Ca. The corresponding state is also visible on the strength of $^{62}$Ca
in Fig. \ref{fig:Ca_ISM_GS}. Wile the proton contribution is negligible, the neutron part displays a behavior similar to what was observed for pure neutron modes at higher energy, i.e. a very broad neutron skin oscillation.

%===================================================================================================================
\subsection{Deformed case}
%==========================================================

%Deformed

The appearance of very low energy pairing mode in deformed and neutron-rich nuclei was already discussed in Ref. \cite{yos06} for quadrupole excitations. In the case of monopole excitations in deformed nuclei, a mixing with the quadrupole modes is expected and shall lead to similar results. Since the onset of pairing correlations is likely to reduce the amount of deformation, the latter is fixed during the calculation in order to focus on the effect of pairing only, in the presence of clusterized strutures.

\begin{figure}[!htb]
\centering
\includegraphics[width=1\linewidth]{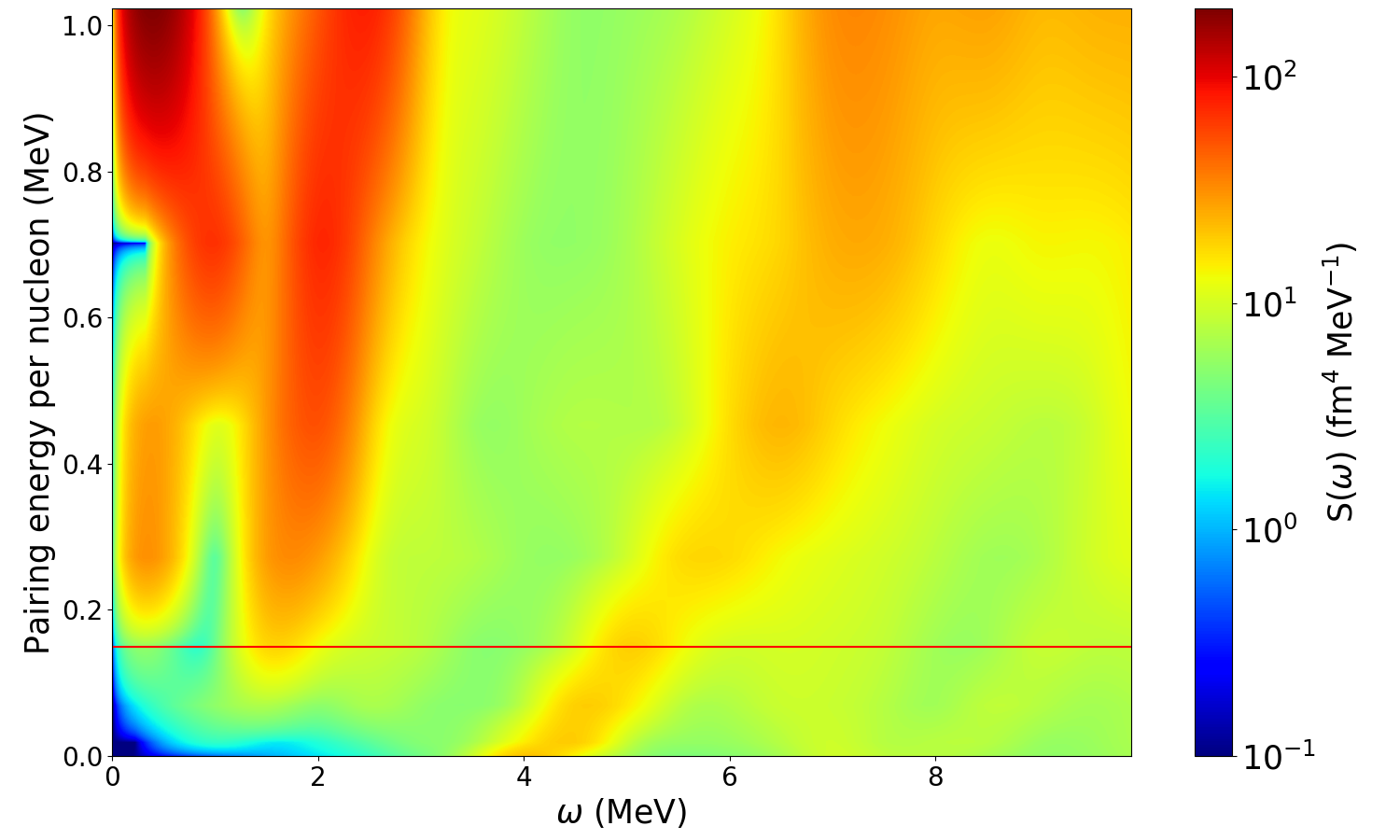}
\caption{(color online). Evolution of the monopole strength function of $^{34}$Mg with the pairing energy of the RHB level. The deformation is fixed to $\beta _2=0.31$ which the GS deformation of this nucleus. The red horizontal line shows the pairing energy of the ground state $^{34}$Mg.}
\label{fig:Mg_pairing}
\end{figure}

Taking $^{34}$Mg as a representative of both deformed and superfluid light nuclei, Fig.\ref{fig:Mg_pairing} displays the evolution of the corresponding ISM strength distribution with the amount of pairing correlations captured by the RHB reference state, with a quadrupole deformation parameter fixed to the ground-state value ($\beta_2=0.31$).

%............................................................
%...........................................................

At zero pairing energy, already few peaks are visible, mainly around 4 MeV. This excitation corresponds to the previously studied cluster excitation, where proton and neutrons behave coherently, generating a cluster oscillations around a core. Two other cluster excitations, at 7 MeV and 9 MeV, carry much less strength and are barely visible in the figure.

Increasing the pairing energy triggers new resonances at very low energy, below 2 MeV. They corresponds to pure pair excitation. 
Interestingly, these transitions involve both protons and neutrons. A calculation of the free response (without the residual interaction) shows that the proton strength, at these very low energy, is negligible. Hence, the residual interaction plays a important role for the pair excitation in neutron-rich nuclei by involving protons. However, this is not a generic property, since for Calcium isotopic chain for instance, no contribution from protons to the strength was found.

In the case of $^{34}$Mg, the origin of these excitations can be traced back to different orbitals, depending on the pairing intensity. For a total pairing energy below 10 MeV, the low-energy excitations are dominated by the 3/2+(2) and 1/2-(3) single-particle states for neutrons, and the 3/2+(1) single-particle state for protons. For larger pairing energies, the neutron contribution to the pairing modes is dominated by the 1/2-(3) and 3/2-(2) orbitals, while the proton contribution is dominated by the 3/2+(1) and 1/2+(2) orbitals.

\begin{figure}[!htb]
\centering
\includegraphics[width=1\linewidth]{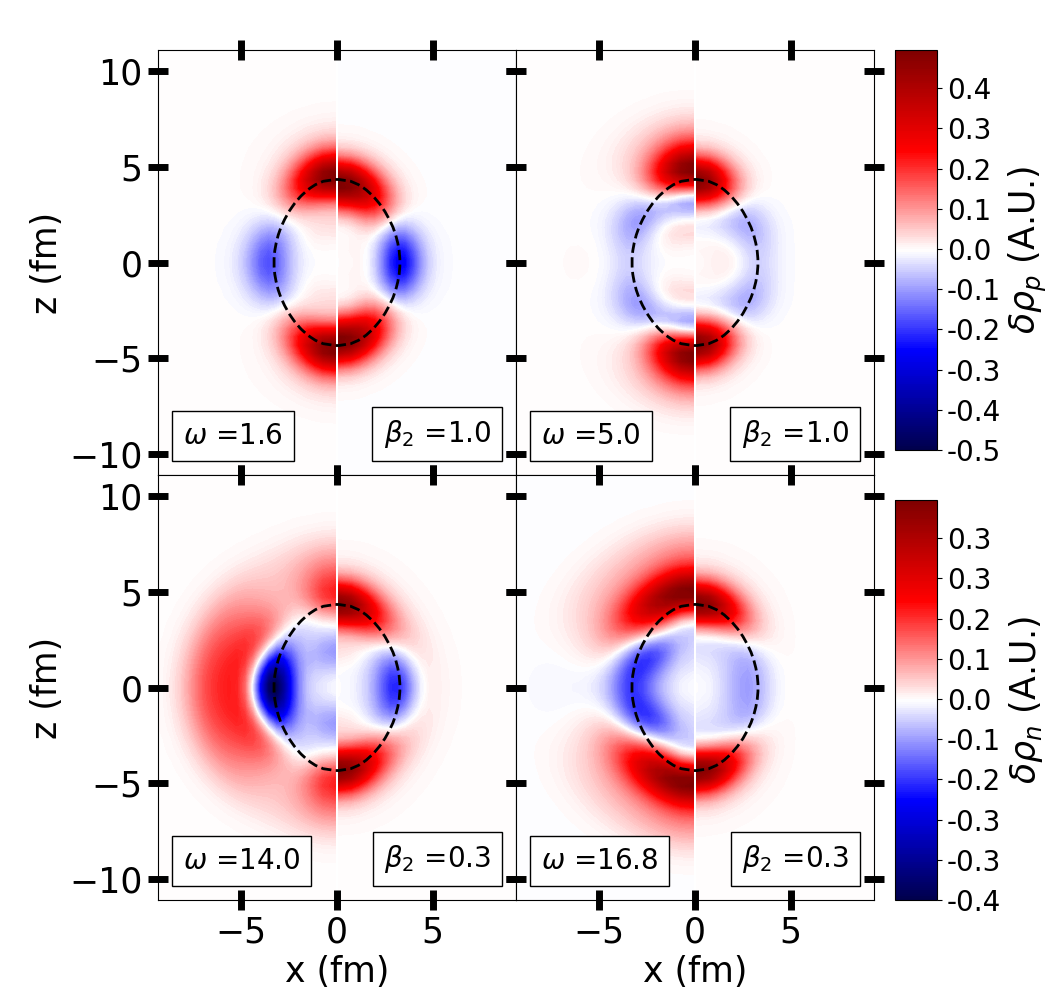}
\caption{(color online). Same as Fig.\ref{fig:Mg_clusters} but for $^{34}$Mg. Excitation energy $\omega$ are given in MeV. The top left plot corresponds to a pairing excitation, and the top right one to a cluster excitation. The bottom left is a pure neutron mode and the bottom right a GR mode.}
\label{fig:Mg_pairing_drho}
\end{figure}

The transition densities of low-energy monopole modes in $^{34}$Mg are displayed in Fig.\ref{fig:Mg_pairing_drho}. Neutrons and protons contribute similarly to the transition density for the two lowest energy modes, i.e. the pairing ($\omega=1.6$ MeV) and the cluster ($\omega=5$ MeV) modes, whereas neutrons dominate the transition density in the $\omega=14$ MeV and $\omega=16.8$ MeV modes, corresponding to the GMR and its low energy tail.

%............................................................
%...........................................................

As for the interplay between pairing and cluster modes, we observe a modification of the structure of the cluster excitation (located at $\omega=5$ MeV in the unconstrained ISM strength) as pairing correlation gets stronger. From the decomposition of the cluster mode into its 2qp component, we observe a transition between a mixture of particle-hole transitions and pair excitation to a pure pair excitation, which dominates because of the partial depletion of the occupied orbitals near the Fermi level.

%
%===================================================================================================================
\section{Conclusion}
%===================================================================================================================
%

\begin{figure}[!htb]
\centering
\includegraphics[width=1\linewidth]{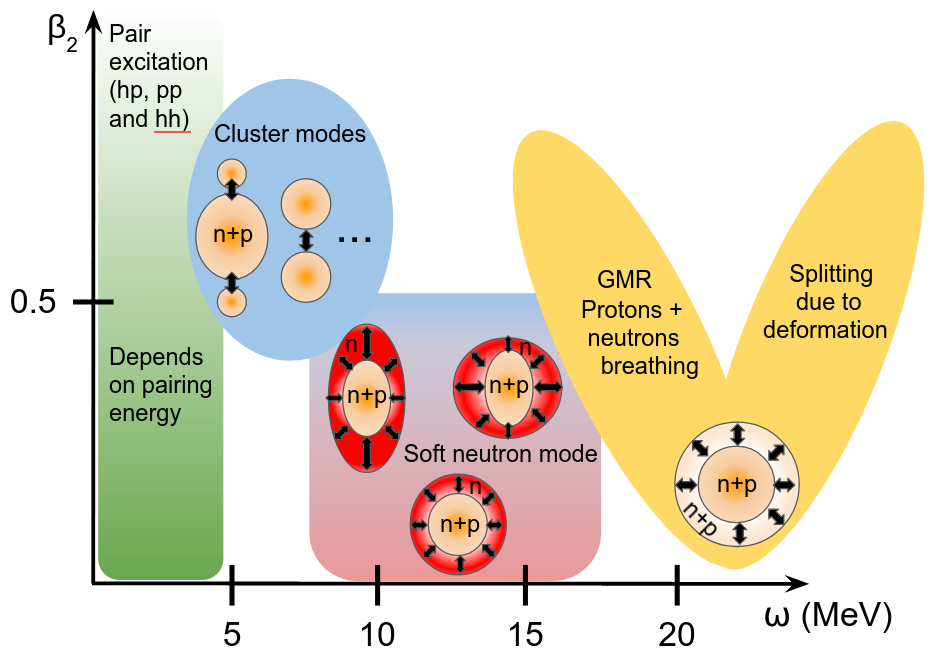}
%\par\end{centering}
\caption{(color online). Schematic view of a typical ISM strength landscape for low mass nuclei (N$<$120). See the main text for more details.}
\label{fig:Sum_up}
\end{figure}

A systematic analysis of the low-lying ISM strength distribution in Neon to Germanium isotopic chains, and especially of the interplay between isospin asymmetry, deformation and superfluidity was performed within a unique microscopic approach, namely the covariant quasi-particle finite amplitude method. The nature and characteristic of the encountered monopole resonances are sketched in~Fig.\ref{fig:Sum_up}.
%............................................................
%...........................................................

Neutron-rich systems feature pure neutron low-energy modes, usually located between 5 to 15 MeV. A detailed analysis of theses modes in terms of 2qp contributions showed that they are built from a single or a couple of single-particle configurations. 

This is explained shell opening effects, due to neutron excess. In particular, it has been shown that magic numbers play an important role: adding two neutrons on top of a magic core, shall lead to the appearance of a new peak at low energy in the monopole strength. Additional appearance of peaks can occur, due to subshell opening. The collective character of the corresponding excitation depends on the number of levels involved in a major shell. These soft neutron modes exhibit a neutron skin, which has been successfully interpreted in terms of the canonical densities, involved in the excitation. 
 
The impact of deformation has been studied by constraining the value quadrupole parameter $\beta _{20}$ over a significant range. This method allows for a better understanding of the evolution of the different excitations with the deformation. In particular, the soft neutron mode remains stable with deformation in most of the cases, and mixes with the GR around $\beta _2 = 0.3$, due to the shift of the GR to lower energy, with deformation. 
Splittings of these soft modes are also visible, and were interpreted as additional excitations, allowed by shell opening, due to the deformation. Different modes were studied and two class of excitations coexist: a mode corresponding to an oscillation along the deformation axis, and another one to an oscillation perpendicular to this one.
Evolving toward very large deformations, destroys the soft neutron mode, and favors the cluster ones.

Finally, pair excitations also emerge below 5 MeV, and correspond to excitation mixing pp, ph and hh channels, inside a single level. These kinds of transition are possible, thanks to pairing, which can modify occupation numbers, allowing for partly occupied levels. The energy related to such an excitation would be $\sim \Delta$ where $\Delta$ refers to the pairing gap. A mixing between pairing and cluster excitations is predicted, the latter transforming into the former, with increasing pairing effects. However, since pairing and deformation are generally competing effects, it is unlikely to find both significant cluster and pairing modes, in the same nucleus.
All these results show that the low energy spectrum of the monopole strength exhibit a rather complex and specific behavior, with respect to neutron excess, deformation and pairing.

\newpage
%===================================================================================================================

\end{document}